\documentclass{article}
\usepackage[outer=25mm,inner=25mm,vmargin=10mm,includehead,includefoot,headheight=15pt]{geometry}
\usepackage{lipsum}
\usepackage{multicol}
\usepackage{graphicx}
\usepackage{caption}
\usepackage{hyperref} 

\usepackage[english]{babel}

\usepackage{colortbl}
\usepackage{color}
\definecolor{pink}{rgb}{1.,0.75,0.8}
\definecolor{green}{rgb}{0.3,1,0.3}
\definecolor{dgreen}{rgb}{0.,0.6,0.}
\definecolor{gold}{rgb}{1.,0.84,0.}
\definecolor{beige}{rgb}{0.96,0.96,0.86}
\definecolor{myyellow}{rgb}{1.,0.84,0.8}
\definecolor{grey}{rgb}{0.8,0.8,0.8}
\definecolor{darkyellow}{cmyk}{0,0,0.5,0.5}
\definecolor{darkwhite}{gray}{0.1}
\definecolor{lightblack}{gray}{0.9}
\definecolor{lightblue}{rgb}{0.1,0.4,1.0}
\definecolor{fucsia}{rgb}{1.,0.4,0.9}

\newenvironment{Figure}
  {\par\medskip\noindent\minipage{\linewidth}}
  {\endminipage\par\medskip}

\begin{document}

\title{Study of 14.1~MeV Neutron Moderation in Beryllium}
	\author{Serena Fattori\footnote{Email: \href{mailto:dr.serena.fattori@gmail.com}{dr.serena.fattori@gmail.com}, ORCID: \href{https://orcid.org/0000-0002-9381-7620}{0000-0002-9381-7620}} , Rino Persiani\thanks{Email: \href{mailto:rinopersiani@gmail.com}{rinopersiani@gmail.com}, ORCID: \href{https://orcid.org/0000-0002-3100-1466}{0000-0002-3100-1466}}, Ugo Abundo}
\date{28 Febbraio 2017}
\maketitle

\begin{abstract}
This study investigates the moderation of 14.1 MeV neutrons in a natural beryllium moderator arranged in a spherical geometry. The neutron interactions and moderation efficiency were analyzed using Monte Carlo simulations with the GEANT4 toolkit. Various sphere radii were tested to determine the optimal moderator thickness for neutron thermalization.
\end{abstract}

\begin{multicols}{2}

\section{Introduction} \label{sec:INT}
This work aims to study a neutron moderator composed of natural beryllium in a spherical geometry, with isotropic 14.1 MeV neutrons generated at the center. The goal is to evaluate neutron moderation across different thicknesses. The geometry was analyzed for thirteen different radii to identify the most effective configuration.

\section{Materials and Methods} \label{sec:MM}
The study of neutron interactions in beryllium and the optimization of the moderator model were carried out using Monte Carlo simulations based on GEANT4 \cite{GEANT4}, version 10.2.

\subsection{Geometry}
The moderator geometry consists of a spherical configuration with a point-like neutron source at the center. Thirteen spheres of different radii were considered: 2.5~cm, 5.0~cm, 7.5~cm, 10.0~cm, 12.5~cm, 15.0~cm, 17.5~cm, 20.0~cm, 25.0~cm, 30.0~cm, 35.0~cm, 40.0~cm, and 45.0~cm.\\

Figure \ref{3Sfere} illustrates three types of events:
\begin{itemize}
\item The first (small sphere, top left) with a radius of 10.0~cm: the neutron (red track) exits the sphere without interacting;
\item The second (medium sphere, center) with a radius of 20.0~cm: the neutron undergoes multiple interactions, producing secondary neutrons and slowing down. Gamma tracks (green) are also visible;
\item The third (large sphere, bottom right) with a radius of 45.0~cm: the neutron undergoes numerous interactions, and all generated neutrons are absorbed without escaping the moderator.
\end{itemize}
These images exemplify typical interaction scenarios for three different thicknesses but do not represent definitive outcomes in 100\% of cases. Quantitative details for all configurations are provided in the subsequent sections of this report.

\begin{Figure}
\centering
  \includegraphics[width=\textwidth]{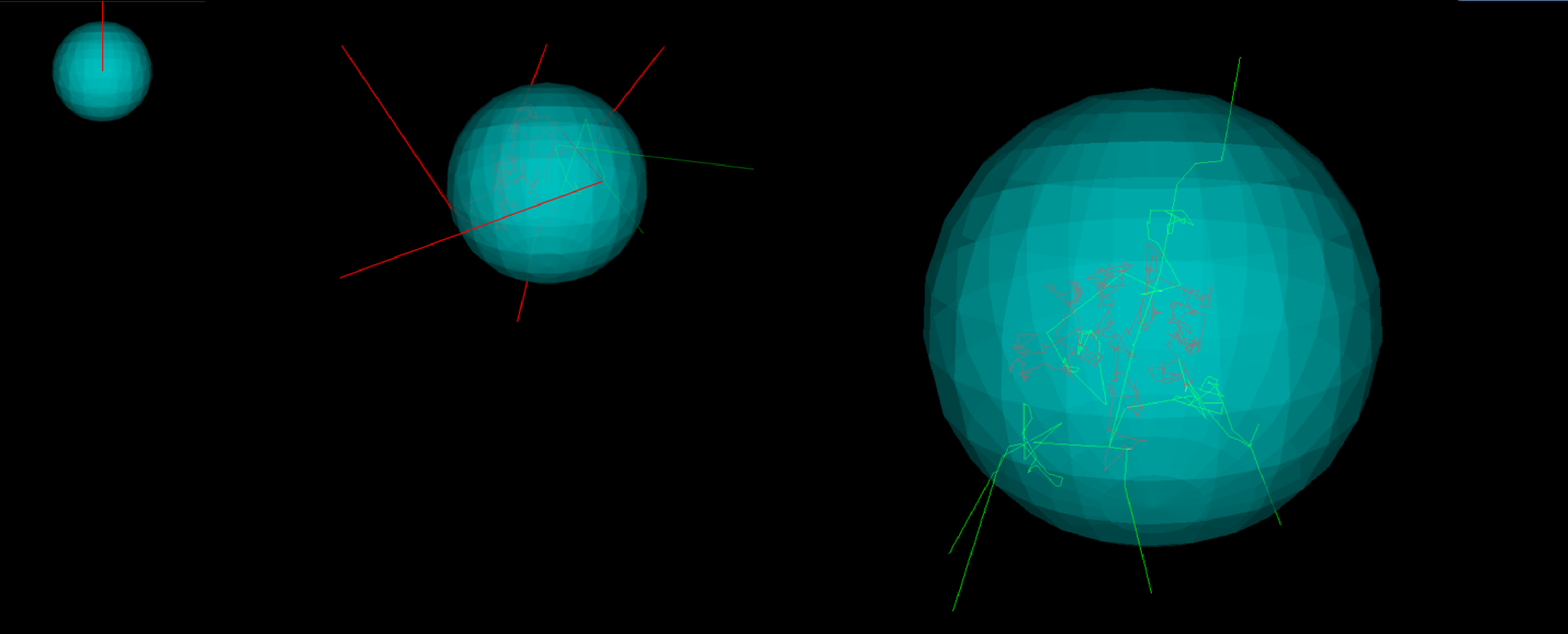}
  \captionof{figure}{Three classic neutron interaction events in a beryllium sphere of: 10.0~cm radius (top left), 20.0~cm radius (center), and 45.0~cm radius (bottom right). Neutron tracks (primary and secondary) are shown in red, while gamma tracks are shown in green.}
  \label{3Sfere}
\end{Figure}

\subsection{Physics}
Various physics lists from the GEANT4 database were examined to compare their performance. The "High Precision" (HP) mode was selected for its accurate description of low-energy processes, though at the cost of slower computation.\\

The primary neutron interactions considered include:
\begin{itemize}
    \item Elastic Scattering;
    \item Inelastic Scattering;
    \item Neutron Capture;
    \item Fission.
\end{itemize}
\noindent Each process is described by several models depending on the energy range. The "High Precision" model was applied for neutron interactions below 20~MeV, utilizing data-driven models. The following models (Figure \ref{Interazioni}) were used \cite{GEANT4models}:
\begin{itemize}
    \item hElasticCHIPS: describes hadron-nucleus elastic scattering using M. Kossov’s parameterized cross-sections;
    \item QGSP: "Quark Gluon String model" for high-energy collisions;
    \item FTFP: describes string formation in hadron-nucleus collisions;
    \item Binary Cascade: models final states in inelastic hadron scattering;
    \item nRadCapture: describes neutron capture at high energy;
    \item G4LFission: models high-energy neutron-induced fission.
\end{itemize}

\begin{Figure}
 \centering
  \includegraphics[width=\linewidth]{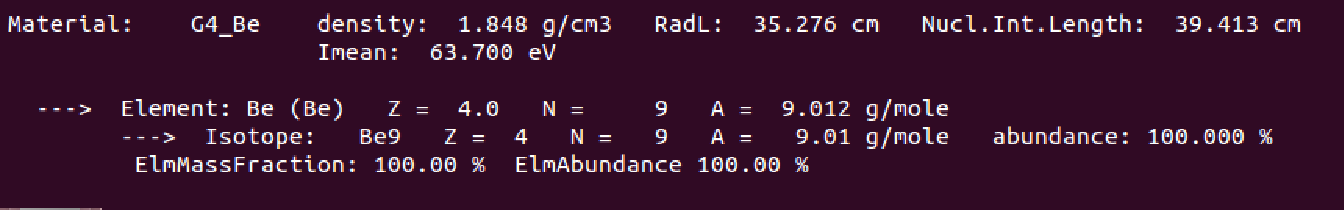}
 \includegraphics[width=\linewidth]{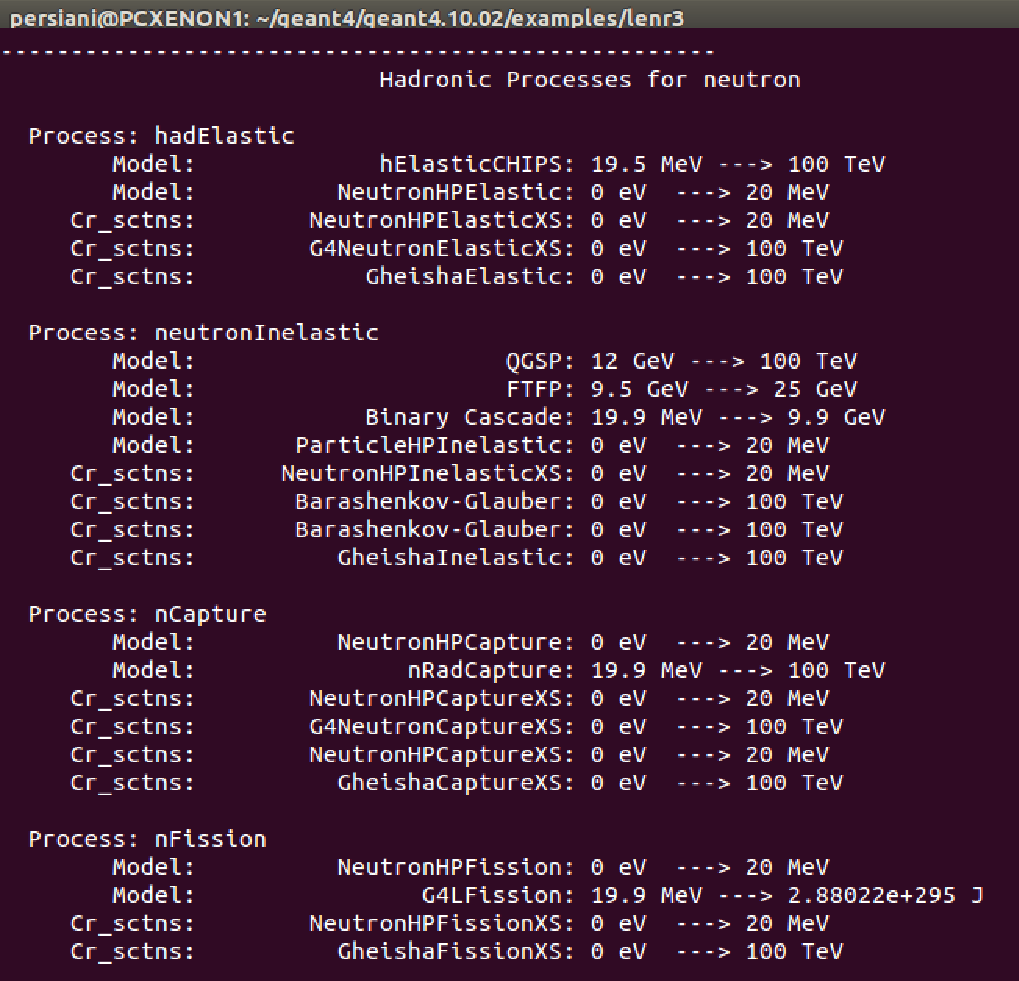}
 \captionof{figure}{Terminal output during simulation execution.}
 \label{Interazioni}
\end{Figure}

\section{Data and Analysis} \label{sec:DA}
For this study, $10^{6}$ neutrons of 14.1~MeV were generated from the center of each of the thirteen spheres.

The following reactions were analyzed:
\begin{itemize}
    \item \begin{equation}
        ^{9}Be + n \to ^{7}Li + t
    \end{equation}
    \item \begin{equation}
        ^{9}Be + n \to n + n + \alpha + \alpha
    \end{equation}
    \item \begin{equation}
        ^{9}Be + n \to  ^{6}He + \alpha
    \end{equation}
\end{itemize}

The output data analyzed include:
\begin{enumerate}
    \item Number of neutrons exiting the sphere;
    \item Number of neutrons with energy $\le 0.025~eV$ exiting the sphere;
    \item Energy spectrum of exiting neutrons;
    \item Energy deposited within the sphere.
\end{enumerate}

\begin{figure*}[hb]
\centering
\hspace{0.5cm}
\includegraphics[width=0.75\linewidth]{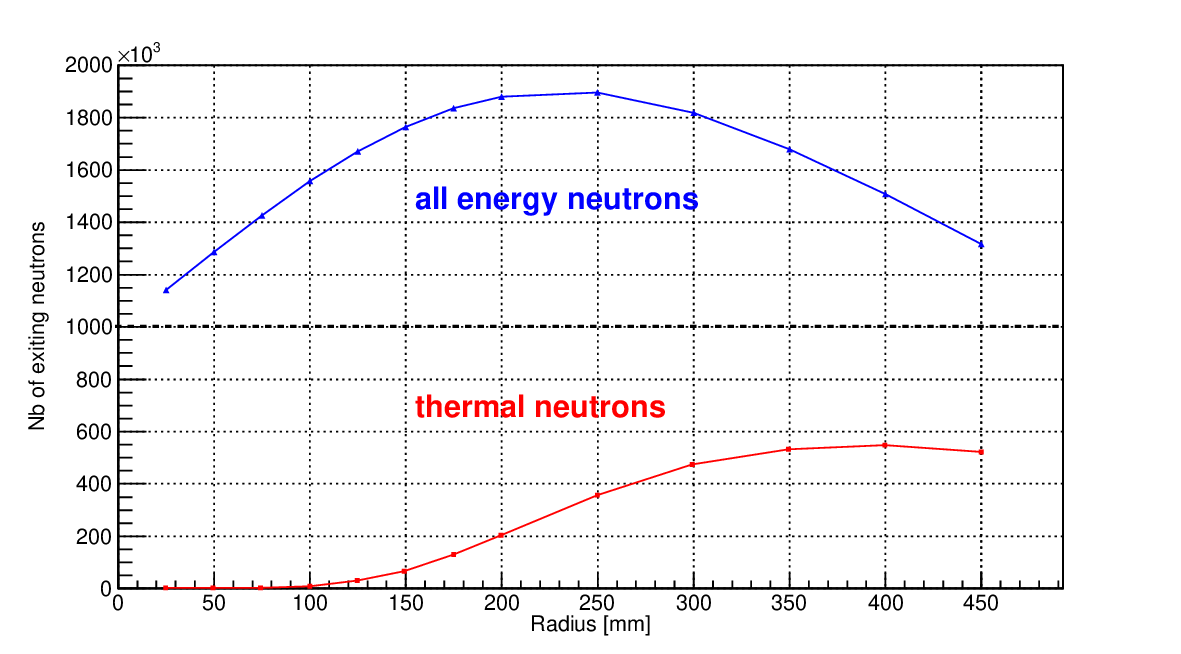}
  \caption{Number of neutrons escaping from the beryllium sphere as a function of its radius: blue represents all energies, while red indicates energies below 0.025~eV. In every case, $10^{6}$ primary events were generated.}
 \label{Nout}
\end{figure*}

\begin{figure*}[p!]
\centering
	\begin{minipage}[t]{0.45\linewidth} 
		\centering
		 \includegraphics[width= 0.90\textwidth]{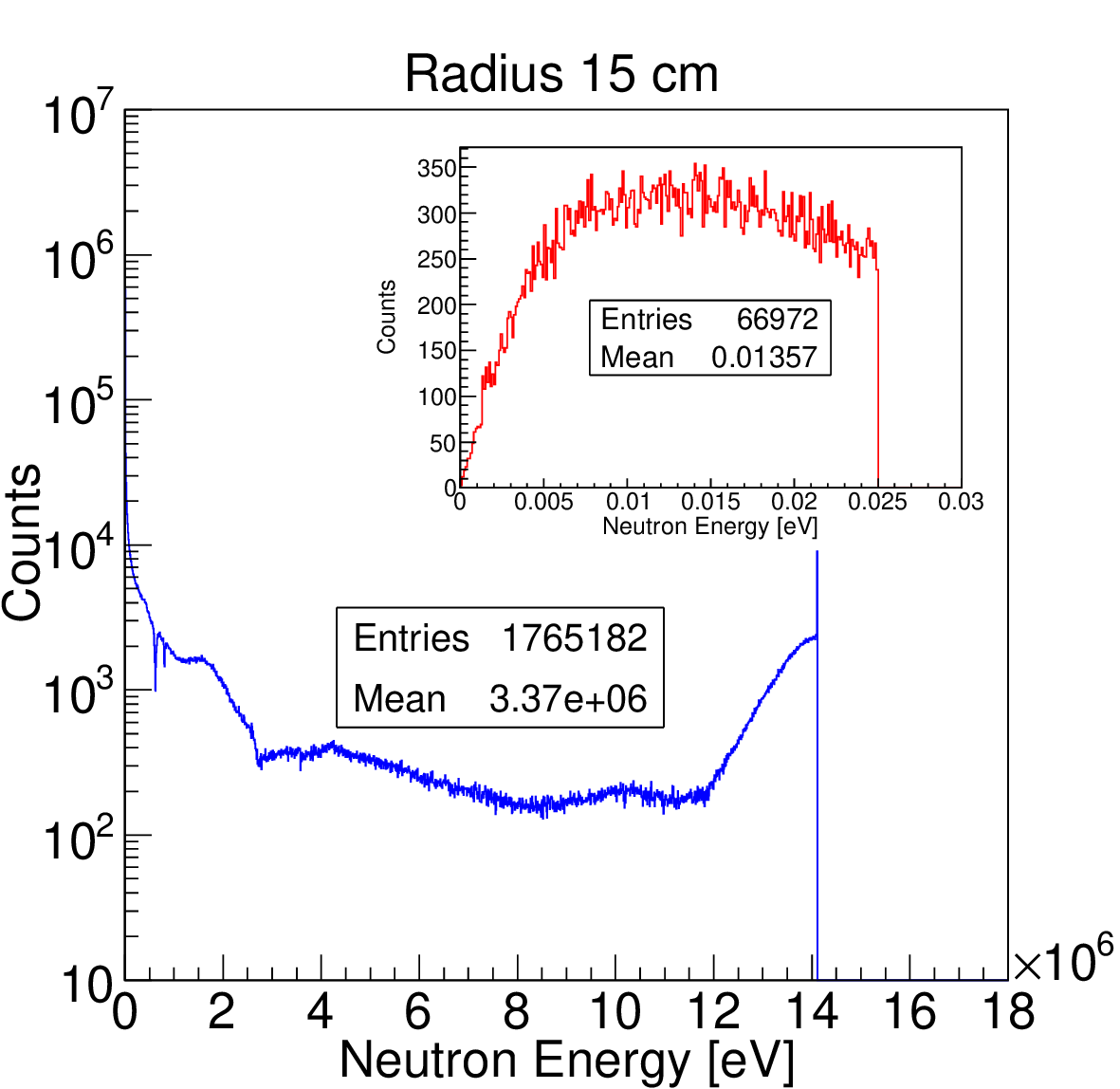} 
	\end{minipage}
		\hspace{0.5cm} 
	\begin{minipage}[t]{0.45\linewidth}
		\includegraphics[width= 0.90\textwidth]{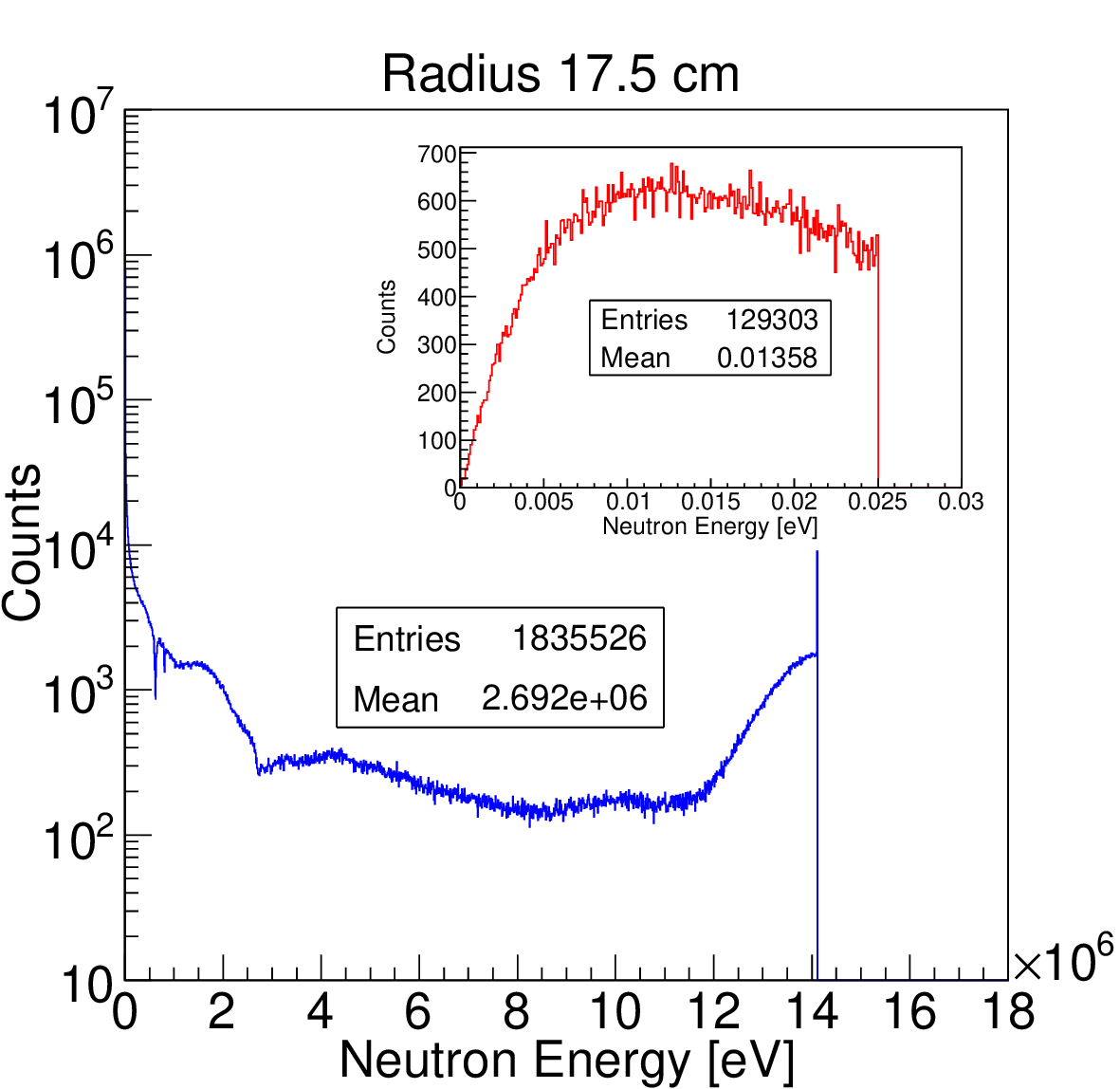} 	 
	\end{minipage}
	\caption{Energy spectra of neutrons escaping from the beryllium sphere for radii of 5.0~cm, 7.5~cm, 10.0~cm, 12.5~cm, 15.0~cm, and 17.5~cm. In each plot, the main blue distribution shows all energies, while the red inset is zoomed in on the energy range corresponding to thermal neutrons ($E_n\le 0.025~eV$).}
	\label{Spettri1}
\end{figure*}

\begin{figure*}[p!]
\centering
	\begin{minipage}[t]{0.45\linewidth} 
		\centering
		 \includegraphics[width= 0.9\textwidth]{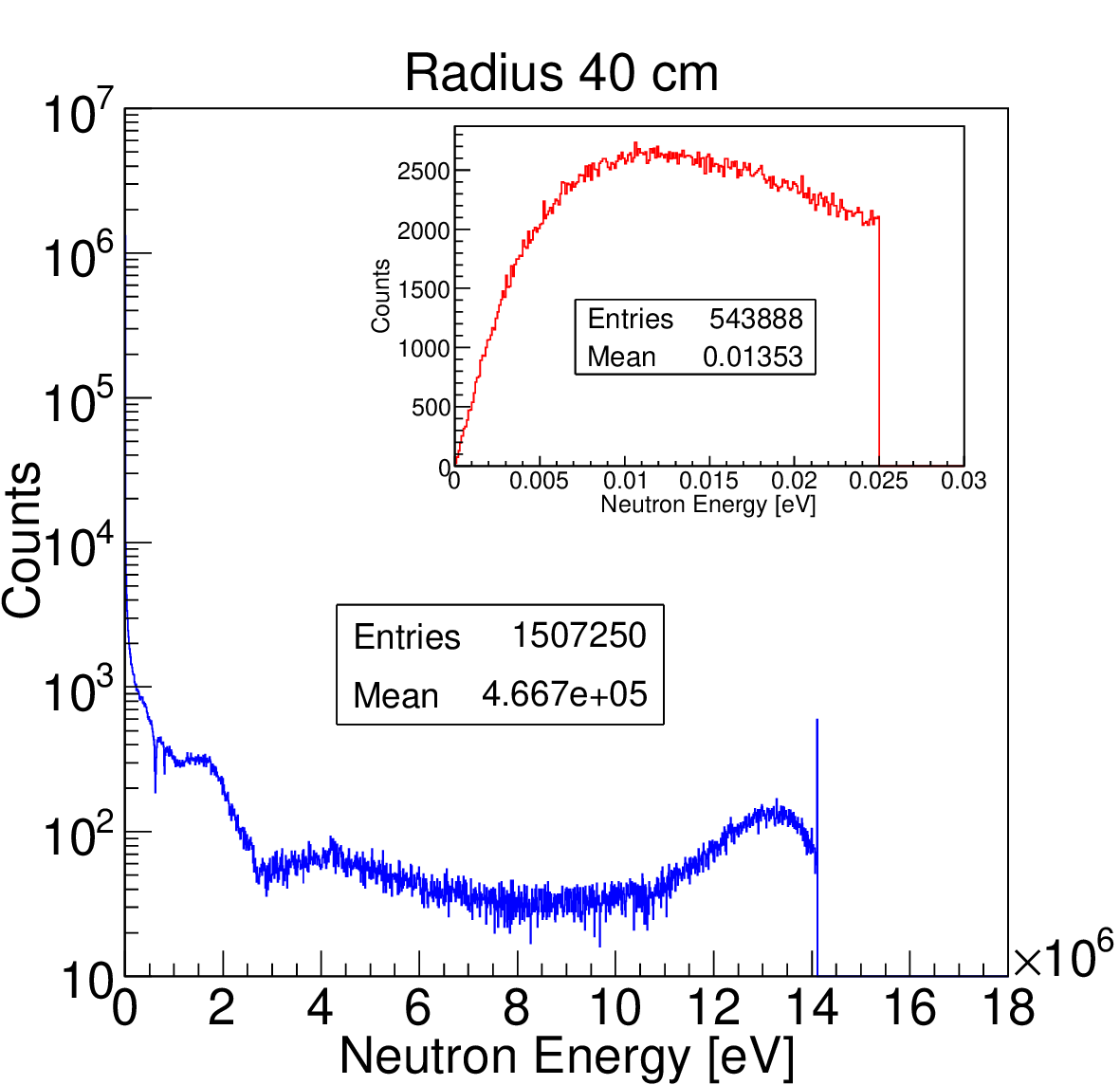} 
	\end{minipage}
		\hspace{0.5cm} 
	\begin{minipage}[t]{0.45\linewidth}
		\includegraphics[width= 0.9\textwidth]{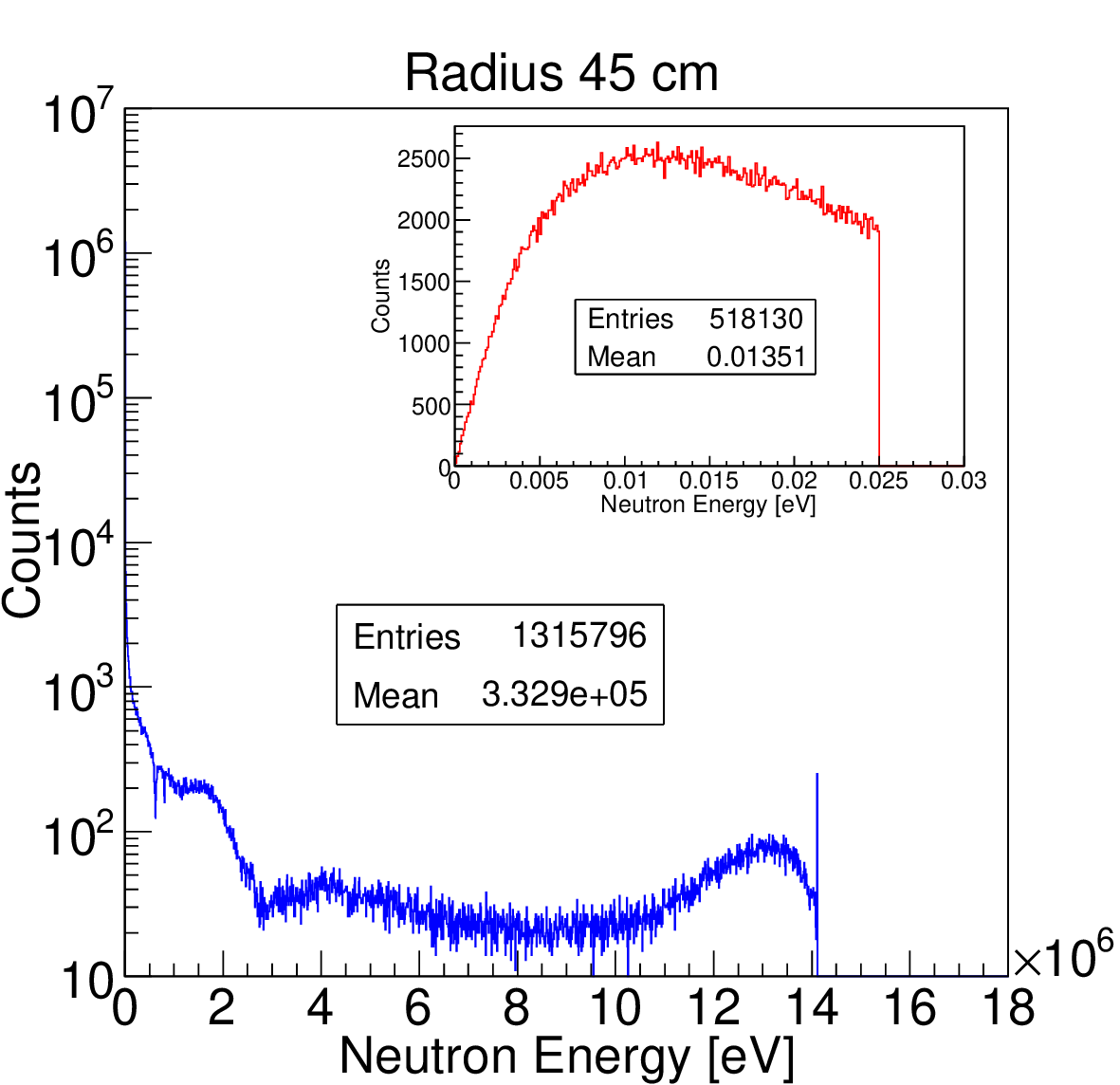} 	 
	\end{minipage}
	\caption{Energy spectra of neutrons escaping from the beryllium sphere for radii of 20.0~cm, 25.0~cm, 30.0~cm, 35.0~cm, 40.0~cm, 45.0~cm. In each plot, the main blue distribution shows all energies, while the red inset is zoomed in on the energy range corresponding to thermal neutrons ($E_n\le 0.025~eV$).}
\label{Spettri2}
\end{figure*}

\begin{figure*}[p!]
\centering
	\begin{minipage}[t]{0.45\linewidth} 
		\centering
		 \includegraphics[width= 0.90\textwidth]{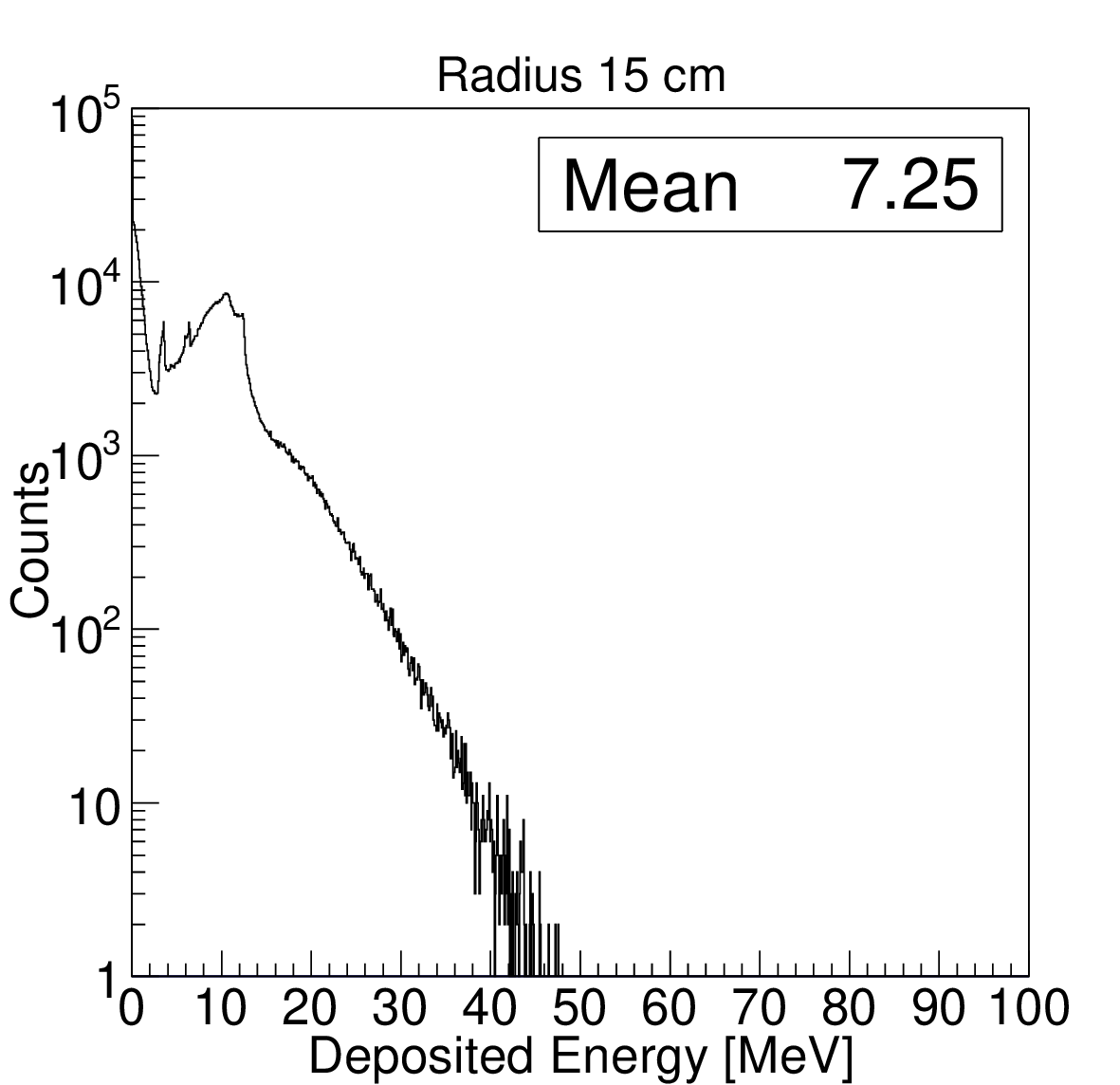} 
	\end{minipage}
		\hspace{0.5cm} 
	\begin{minipage}[t]{0.45\linewidth}
		\includegraphics[width= 0.90\textwidth]{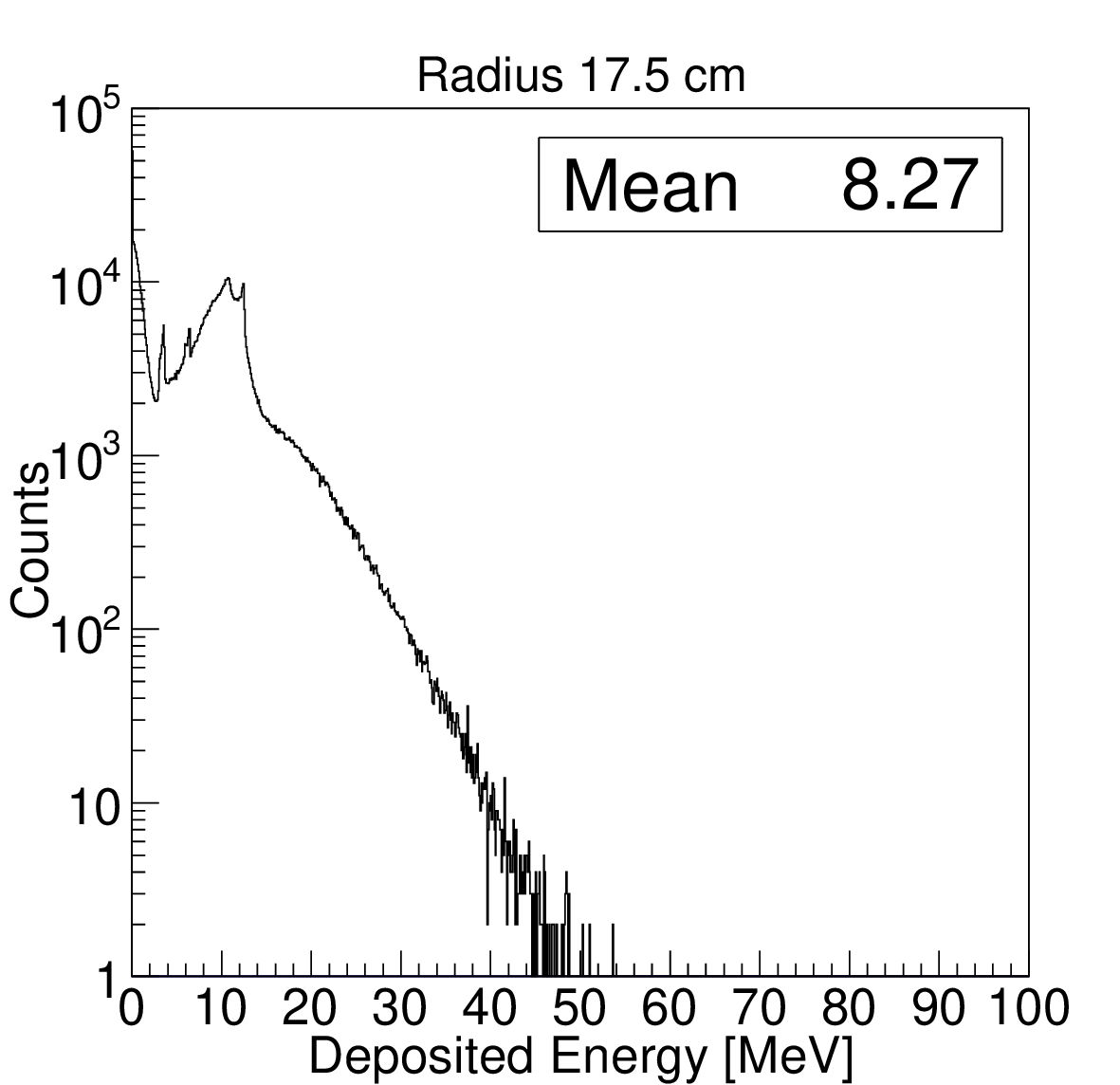} 	 
	\end{minipage}
	\caption{Energy deposition spectra of 14.1~MeV neutrons in the beryllium sphere for radii of 5.0~cm, 7.5~cm, 10.0~cm, 12.5~cm, 15.0~cm, 17.5~cm.}
	\label{Spettri3}
\end{figure*}

\begin{figure*}[p!]
\centering
	\begin{minipage}[t]{0.45\linewidth} 
		\centering
		 \includegraphics[width= 0.9\textwidth]{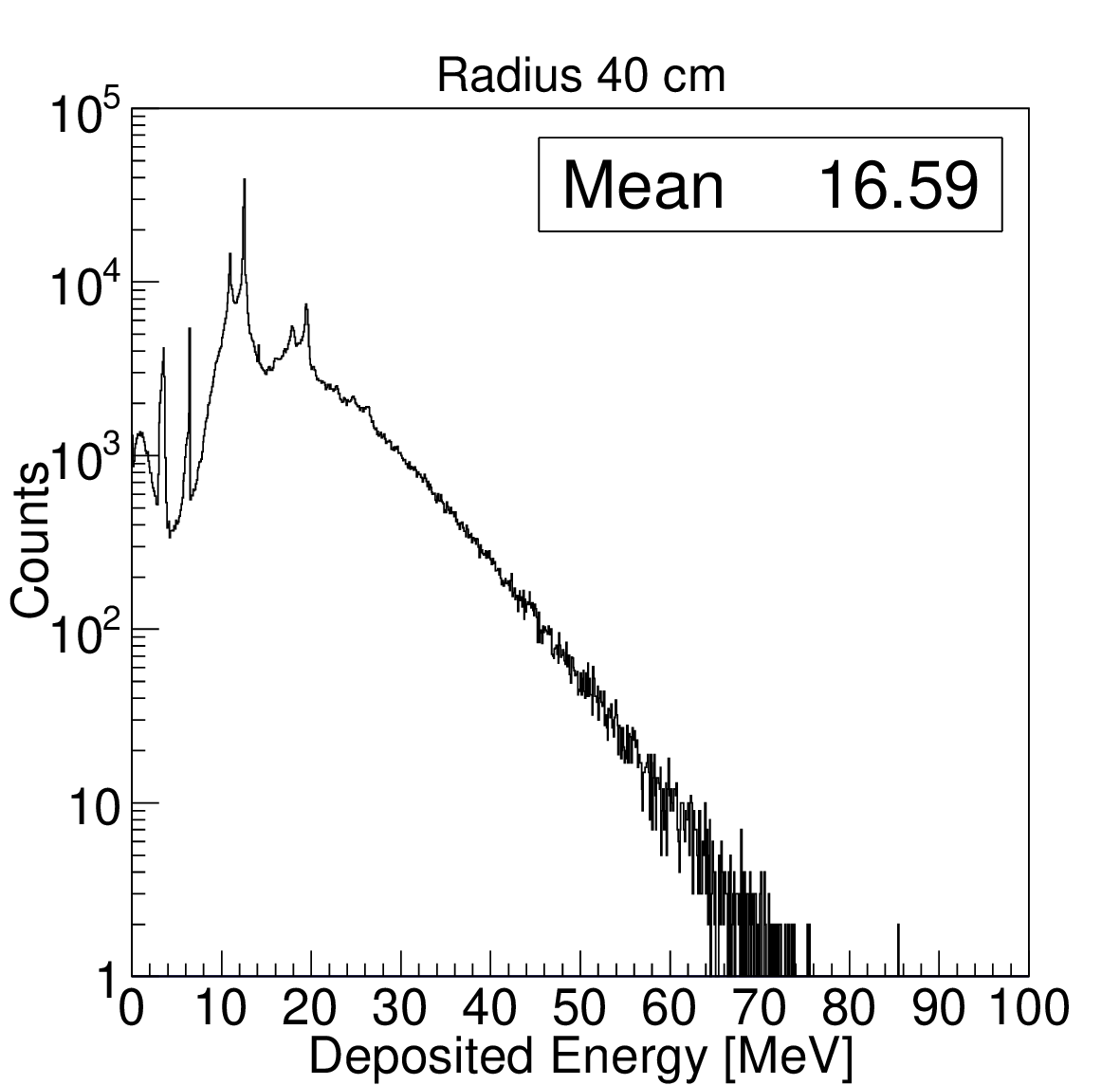} 
	\end{minipage}
		\hspace{0.5cm} 
	\begin{minipage}[t]{0.45\linewidth}
		\includegraphics[width= 0.9\textwidth]{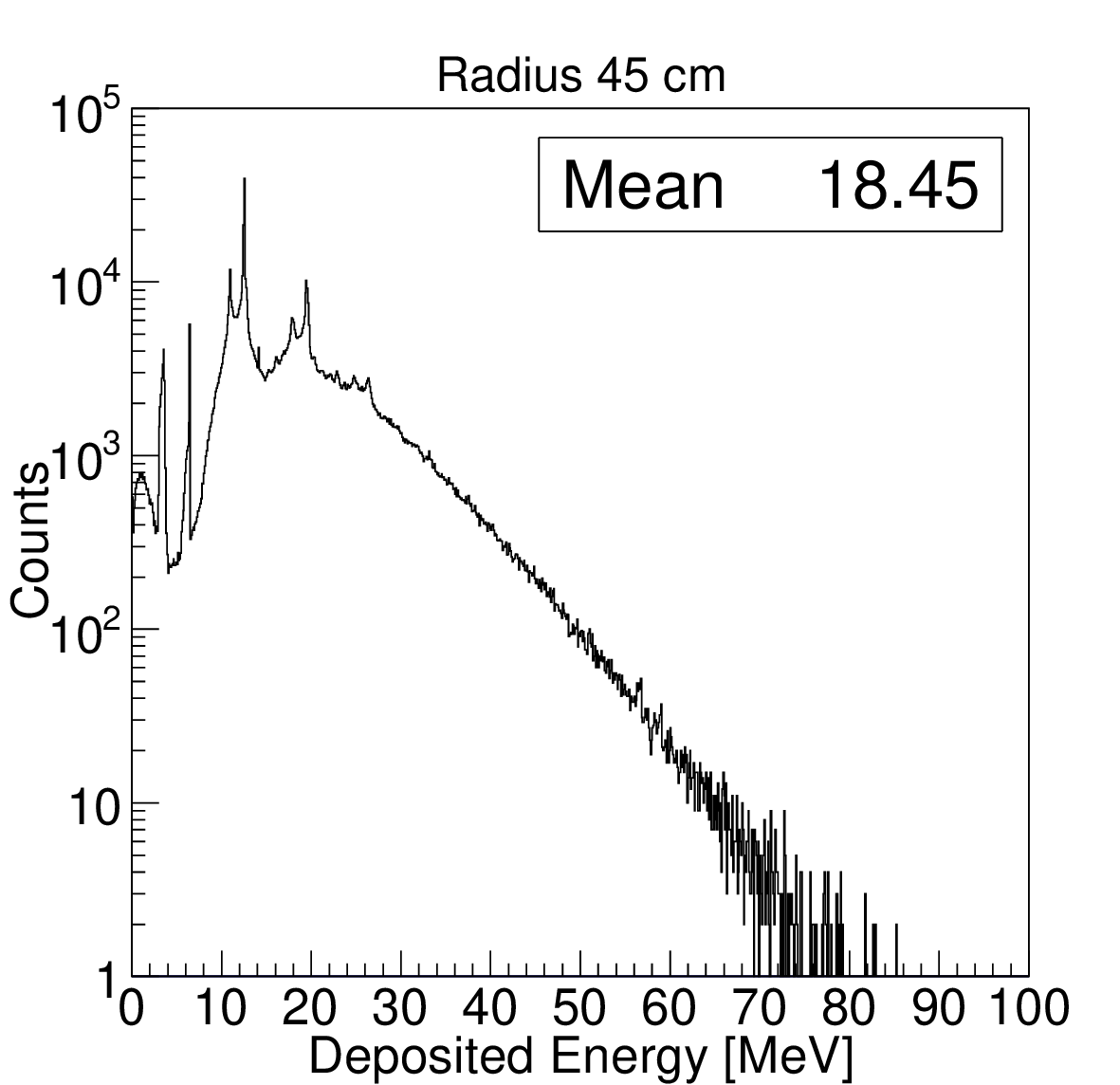} 	 
	\end{minipage}
	\caption{Energy deposition spectra of 14.1~MeV neutrons in the beryllium sphere for radii of 20.0~cm, 25.0~cm, 30.0~cm, 35.0~cm, 40.0~cm, 45.0~cm.}
\label{Spettri4}
\end{figure*}

The results for the first two points are presented in Table \ref{tabella_Nneutroni} and Figure \ref{Nout}. Energy spectra are shown in Figures \ref{Spettri1} and \ref{Spettri2}, while deposited energy distributions are illustrated in Figures \ref{Spettri3} and \ref{Spettri4}.

\section{Conclusions} \label{sec:CON}
 Based on the results obtained (Table \ref{tabella_Nneutroni} and Figures \ref{Spettri1}, \ref{Spettri2}, \ref{Spettri3}, \ref{Spettri4}), we see that the yield of thermal neutrons is negligible for moderator thicknesses below 10~cm. A significant increase occurs for thicknesses greater than 10~cm, peaking at 40~cm before decreasing at larger thicknesses. The optimal moderator thickness is between 20cm and 30cm, as this range maximizes both total neutron production and thermal neutron output. Neutrons with slightly higher energies could still be moderated in a second step when crossing the Lithium Deuteride fuel cell.
  
 \section*{Acknowledgments} \label{sec:RIN}
 We thank Futureon s.r.l. Centro Ricerche Energetiche, via Acqua Donzella 33, 00179 Rome, for supporting this work.

\end{multicols}

\begin{table}[h!]
\begin{center}
\resizebox{7.7cm}{!}{ 
\begin{tabular}{ |c|c|c| }
\hline
Radius &  Neutrons 				& Neutrons $E_{n} \le 0.025~eV$  \\   
$[cm]$ &  [$\sharp$]  			& [$\sharp$]  \\    
\hline
2.5 &  1142400 $\pm$ 1100		& 0 $\pm$ 0\\
5.0 & 1286500 $\pm$ 1100		& 2	$\pm$ 1\\
7.5 & 1425500 $\pm$ 1200		& 380 $\pm$ 20\\
10.0 & 1556600 $\pm$ 1200		& 5230	$\pm$ 70\\
12.5 & 1670800 $\pm$ 1300		& 25150 $\pm$ 160\\
15.0 & 1765200 $\pm$ 1300		& 67000	$\pm$ 300\\
17.5 & 1835500 $\pm$ 1400		& 129300	$\pm$ 400\\
20.0 & 1879700 $\pm$ 1400		& 204300	$\pm$ 500\\
25.0 & 1894300 $\pm$ 1400		& 357200	$\pm$ 600\\
30.0 & 1817800 $\pm$ 1300		& 472000	$\pm$ 700\\
35.0 & 1678100 $\pm$ 1300		& 531900	$\pm$ 700\\
40.0 & 1507300 $\pm$ 1200		& 543900	$\pm$ 700\\
45.0 & 1315800 $\pm$ 1100		& 518100	$\pm$ 700\\
\hline
\end{tabular}        

} 
\end{center}
  \caption{As the sphere radius (provided in the first column) varies, the number of neutrons escaping the sphere is reported for all energies (second column) and specifically for energies below 0.025~eV (third column). The errors shown represent the statistical uncertainty.}
   \label{tabella_Nneutroni}
\end{table}

\begin{figure*}[p!]
\centering
	\begin{minipage}[t]{0.45\linewidth} 
		\centering
		 \includegraphics[width= 0.90\textwidth]{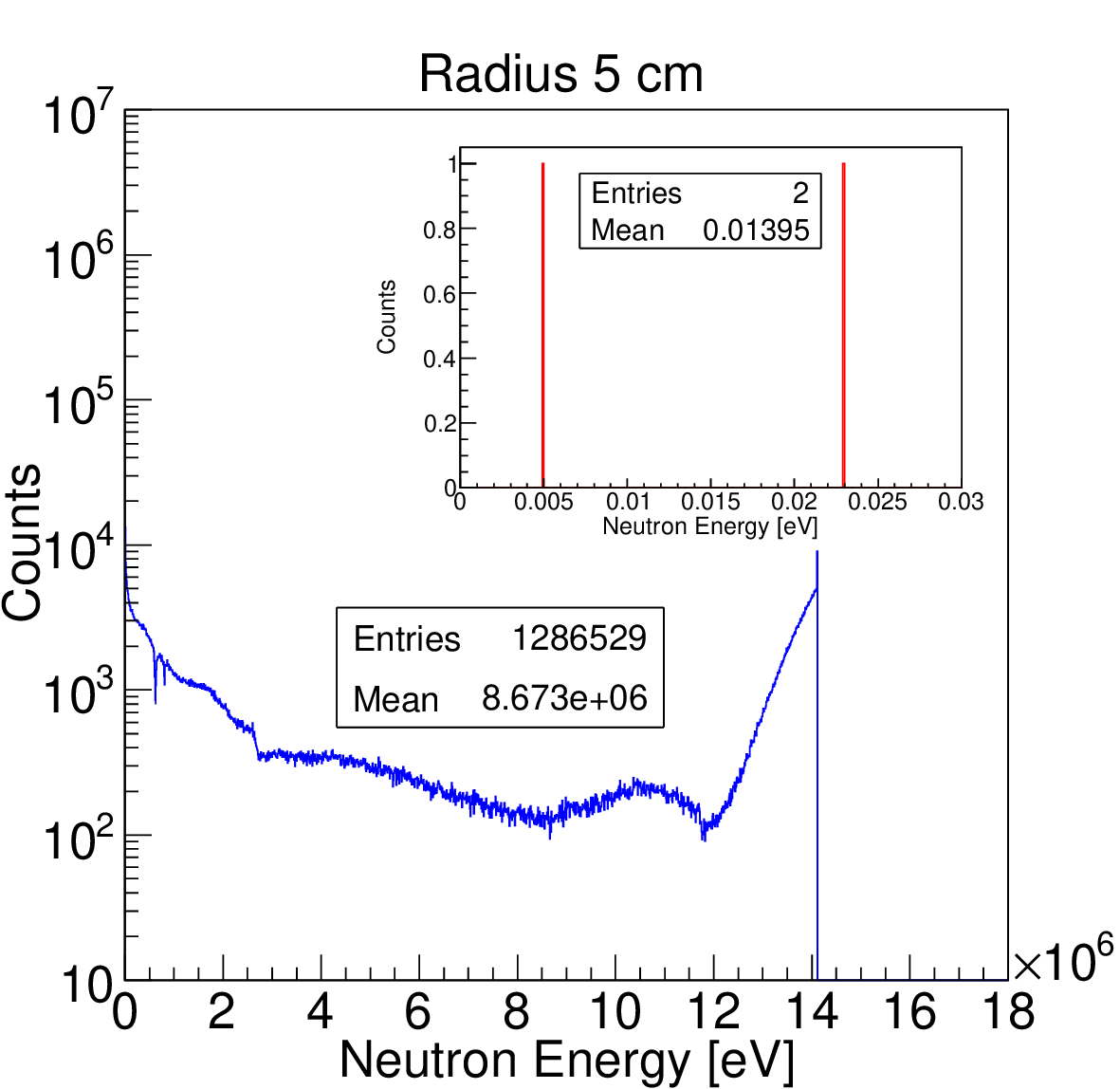} 
	\end{minipage}
		\hspace{0.5cm} 
	\begin{minipage}[t]{0.45\linewidth}
		\includegraphics[width= 0.90\textwidth]{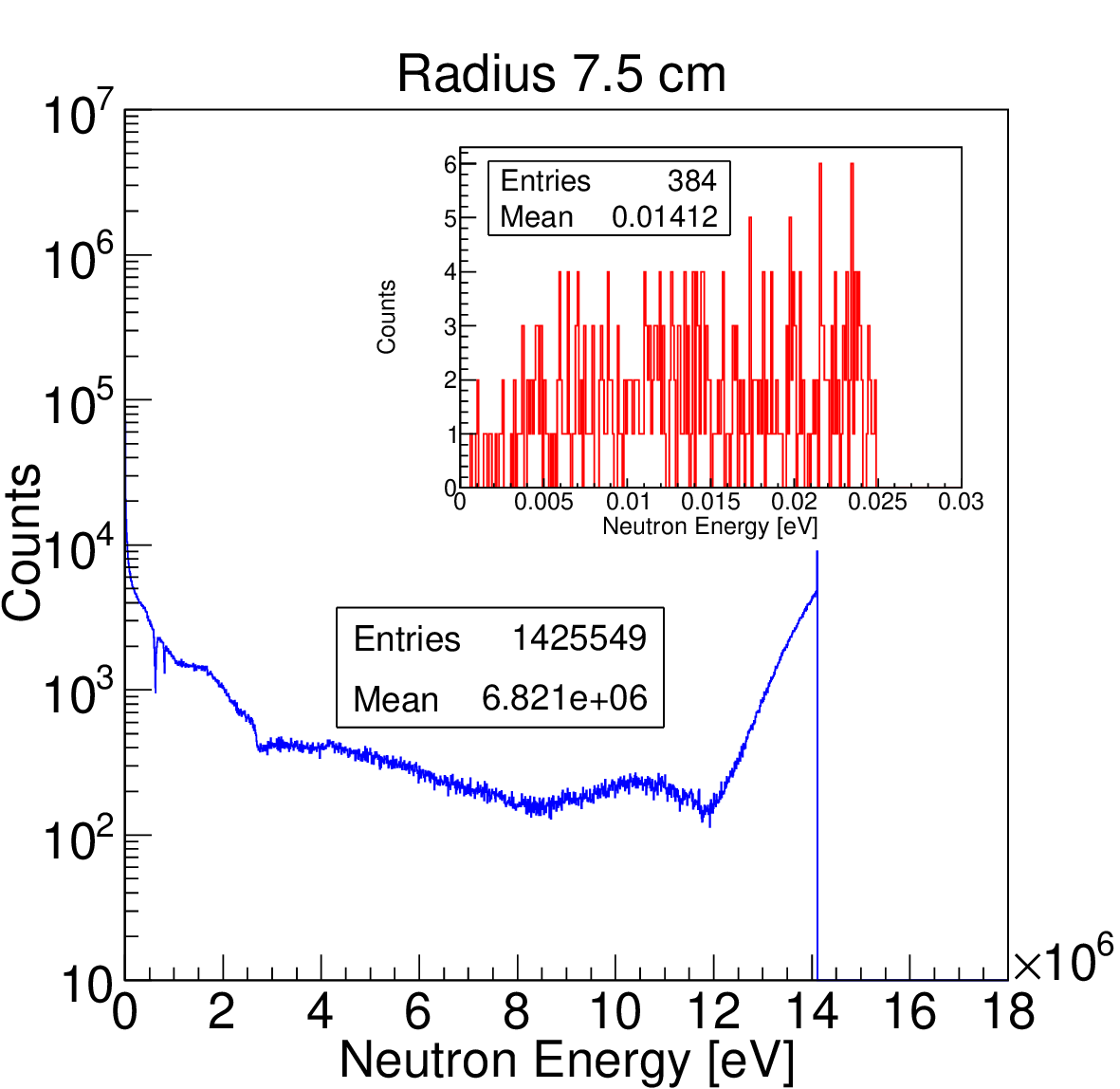} 
	\end{minipage}
\end{figure*}

\begin{figure*}[p!]
\centering
	\begin{minipage}[t]{0.45\linewidth} 
		\centering
		 \includegraphics[width= 0.90\textwidth]{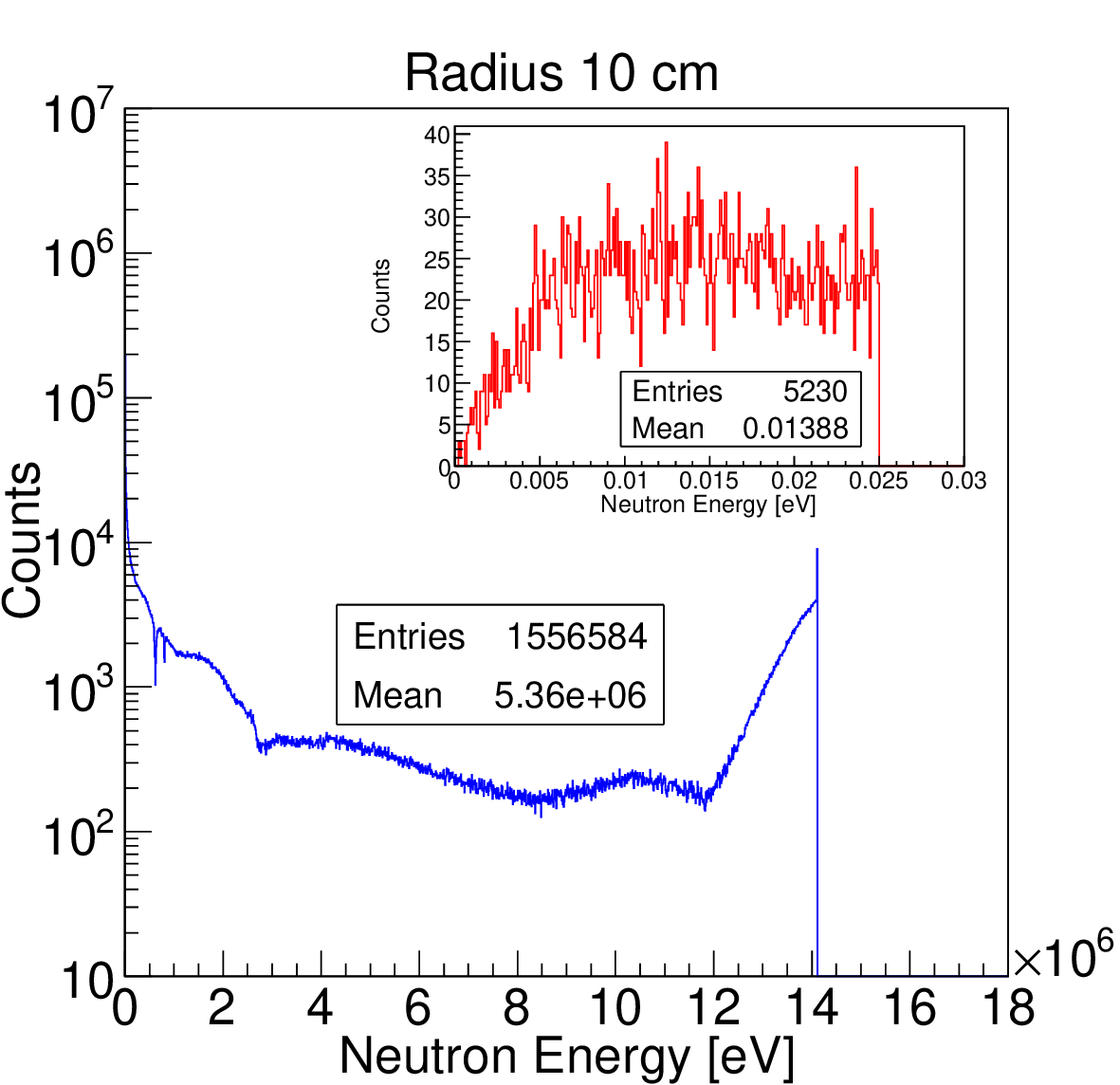} 
	\end{minipage}
		\hspace{0.5cm} 
	\begin{minipage}[t]{0.45\linewidth}
		\includegraphics[width= 0.90\textwidth]{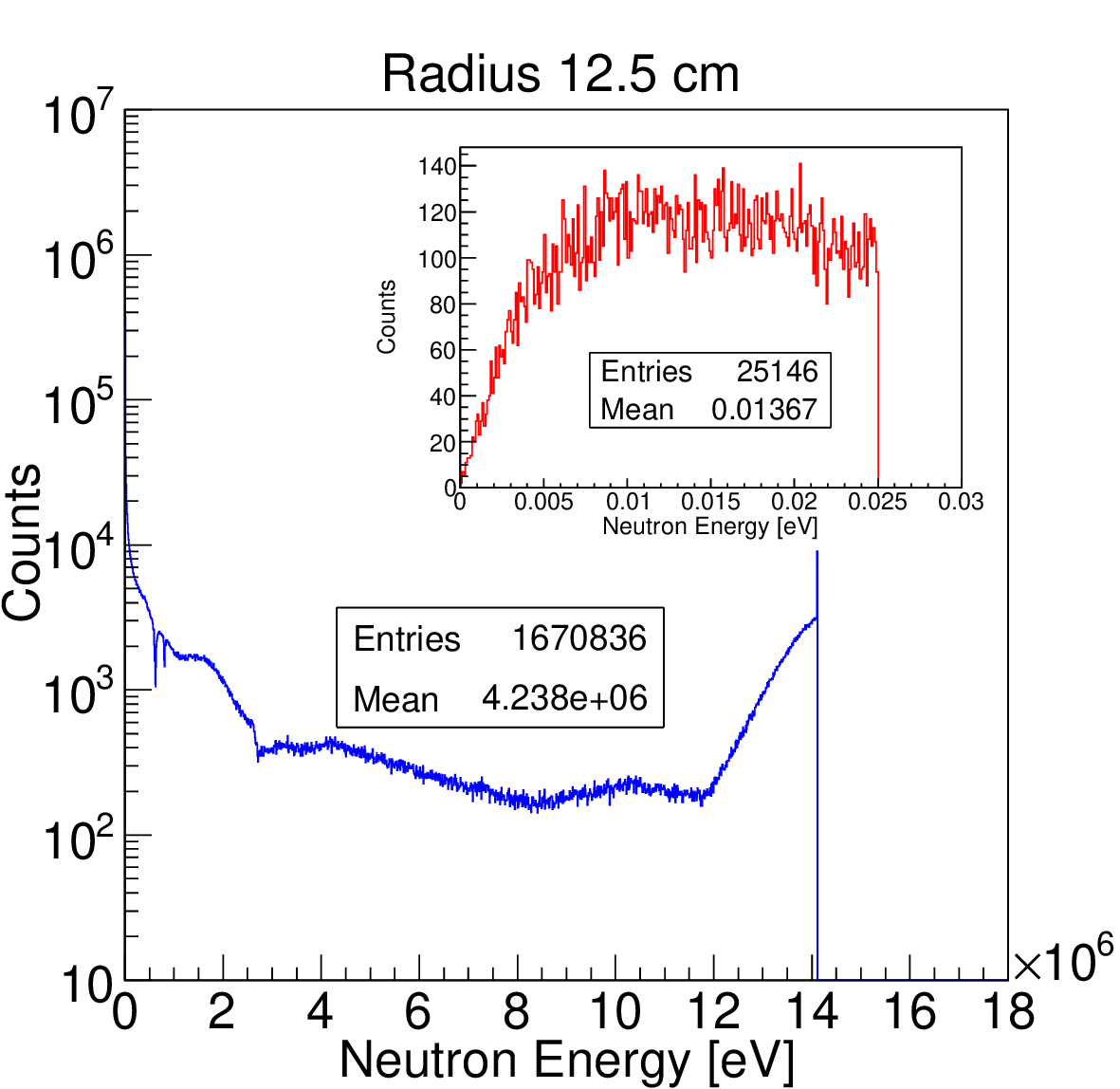} 		 
	\end{minipage}
\end{figure*}

\begin{figure*}[p!]
\centering
	\begin{minipage}[t]{0.45\linewidth} 
		\centering
		 \includegraphics[width= 0.90\textwidth]{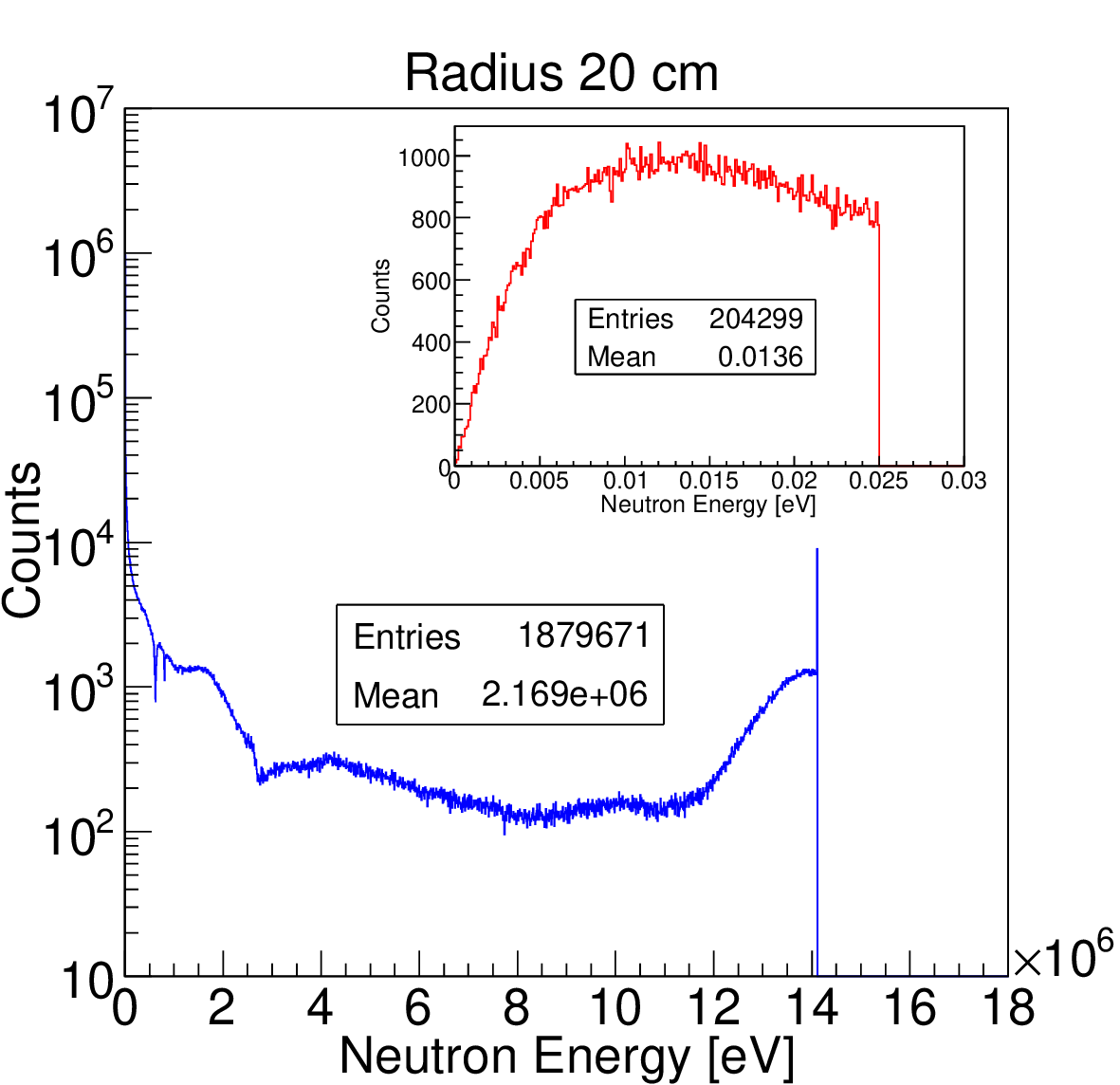} 
	\end{minipage}
		\hspace{0.5cm} 
	\begin{minipage}[t]{0.45\linewidth}
		\includegraphics[width= 0.9\textwidth]{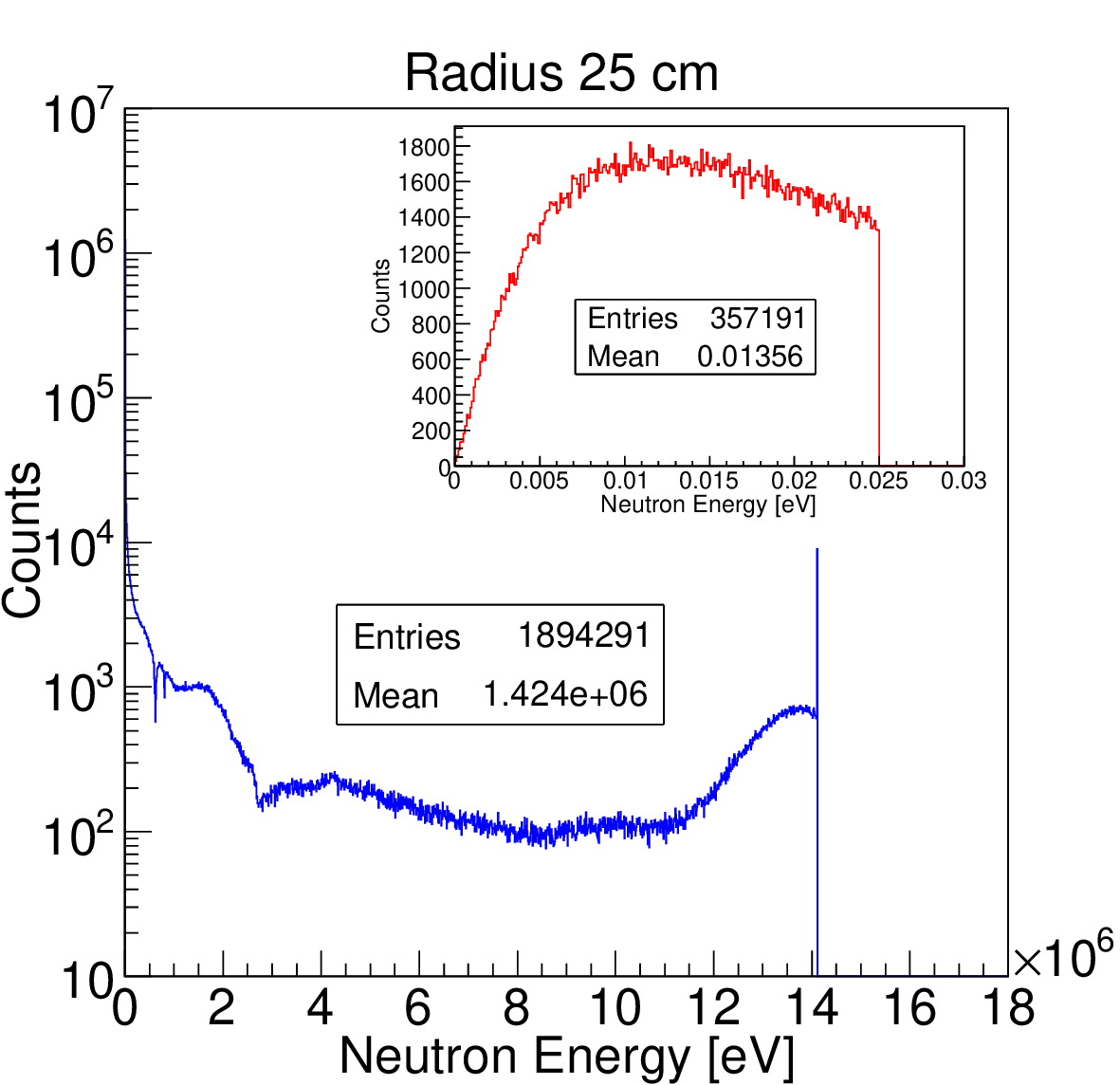} 	 
	\end{minipage}
\end{figure*}

\begin{figure*}[p!]
\centering
	\begin{minipage}[t]{0.45\linewidth} 
		\centering
		 \includegraphics[width= 0.9\textwidth]{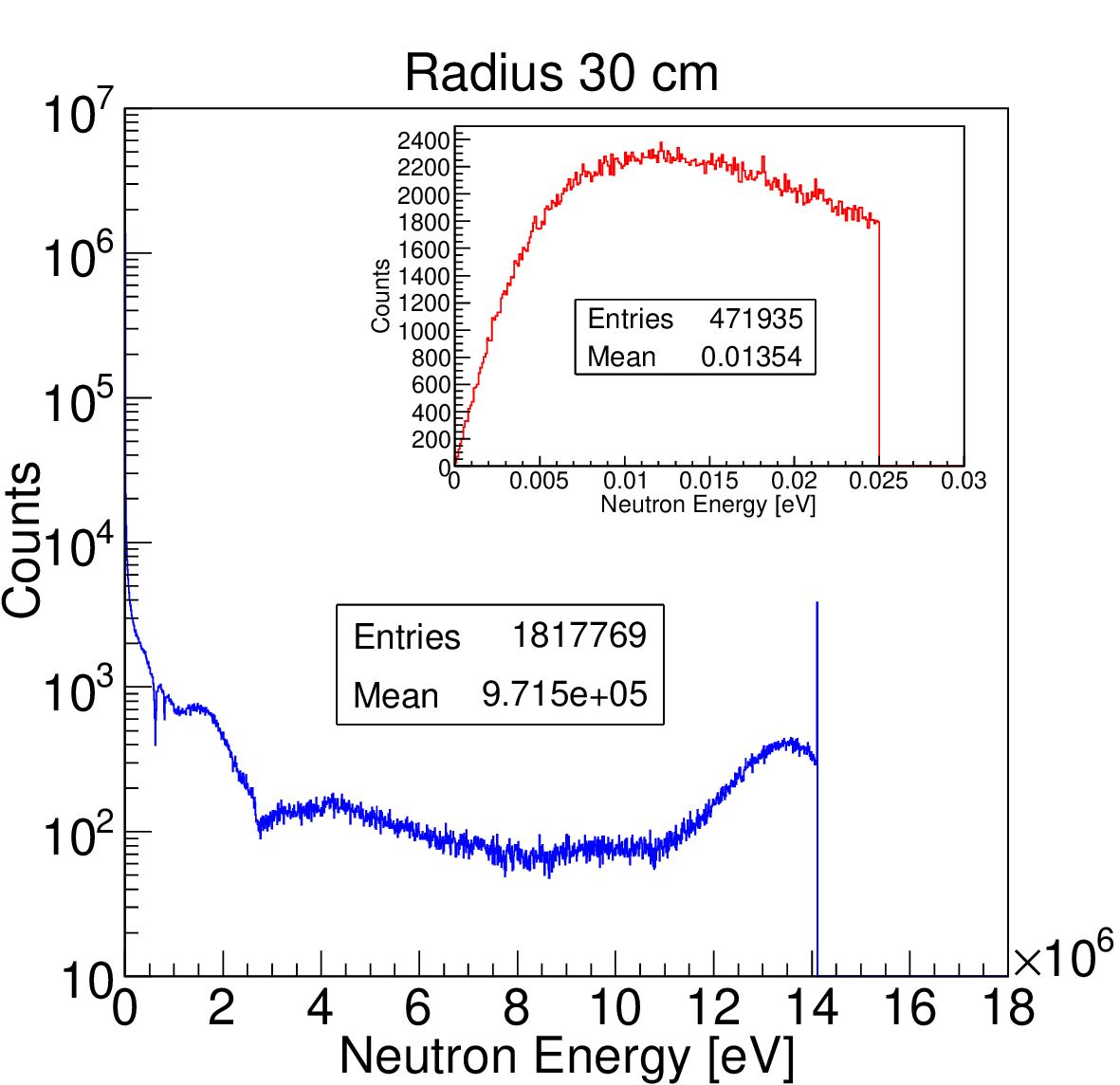} 
	\end{minipage}
		\hspace{0.5cm} 
	\begin{minipage}[t]{0.45\linewidth}
		\includegraphics[width= 0.9\textwidth]{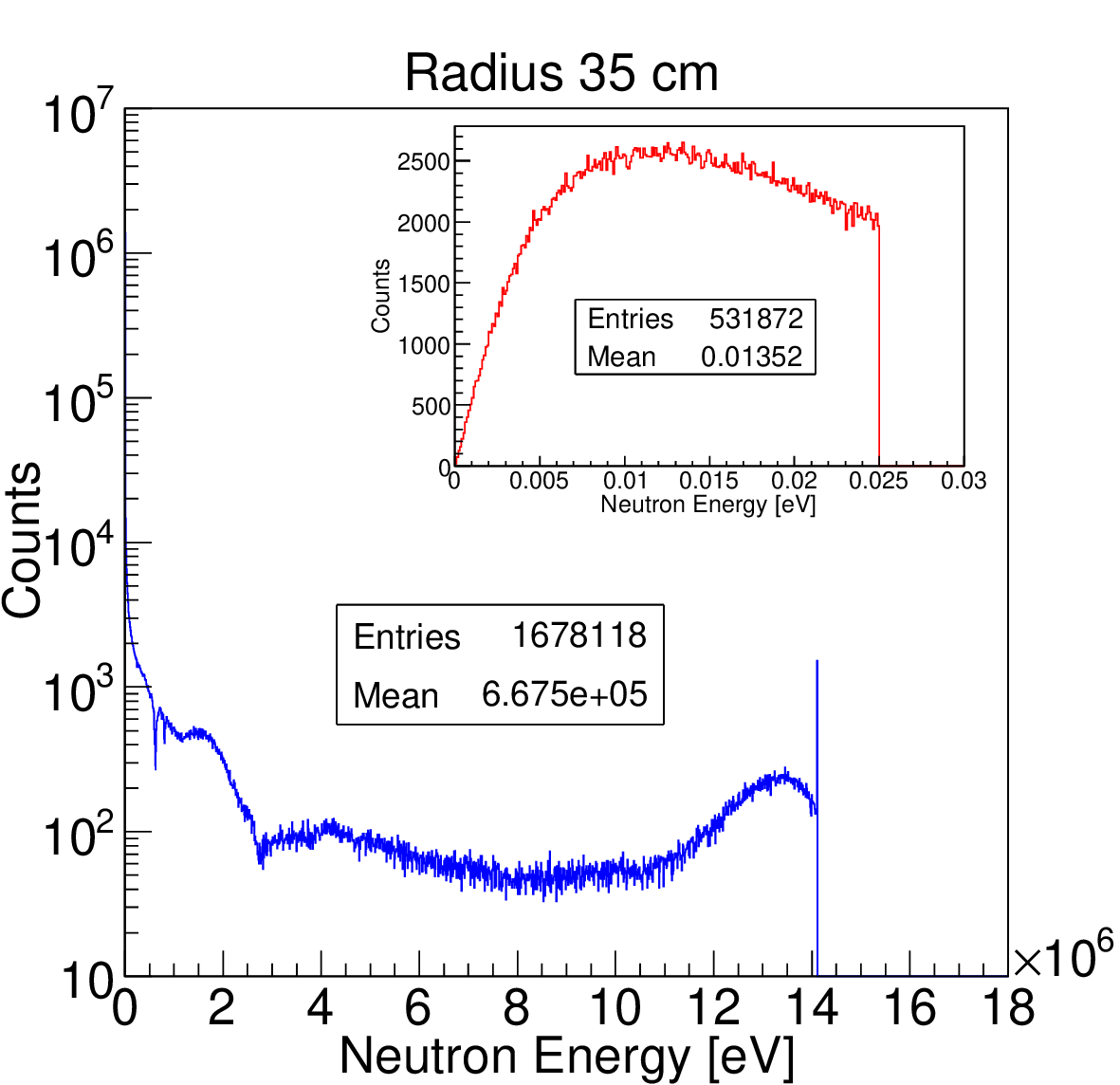} 	 
	\end{minipage}
\end{figure*}

\begin{figure*}[p!]
\centering
	\begin{minipage}[t]{0.45\linewidth} 
		\centering
		 \includegraphics[width= 0.90\textwidth]{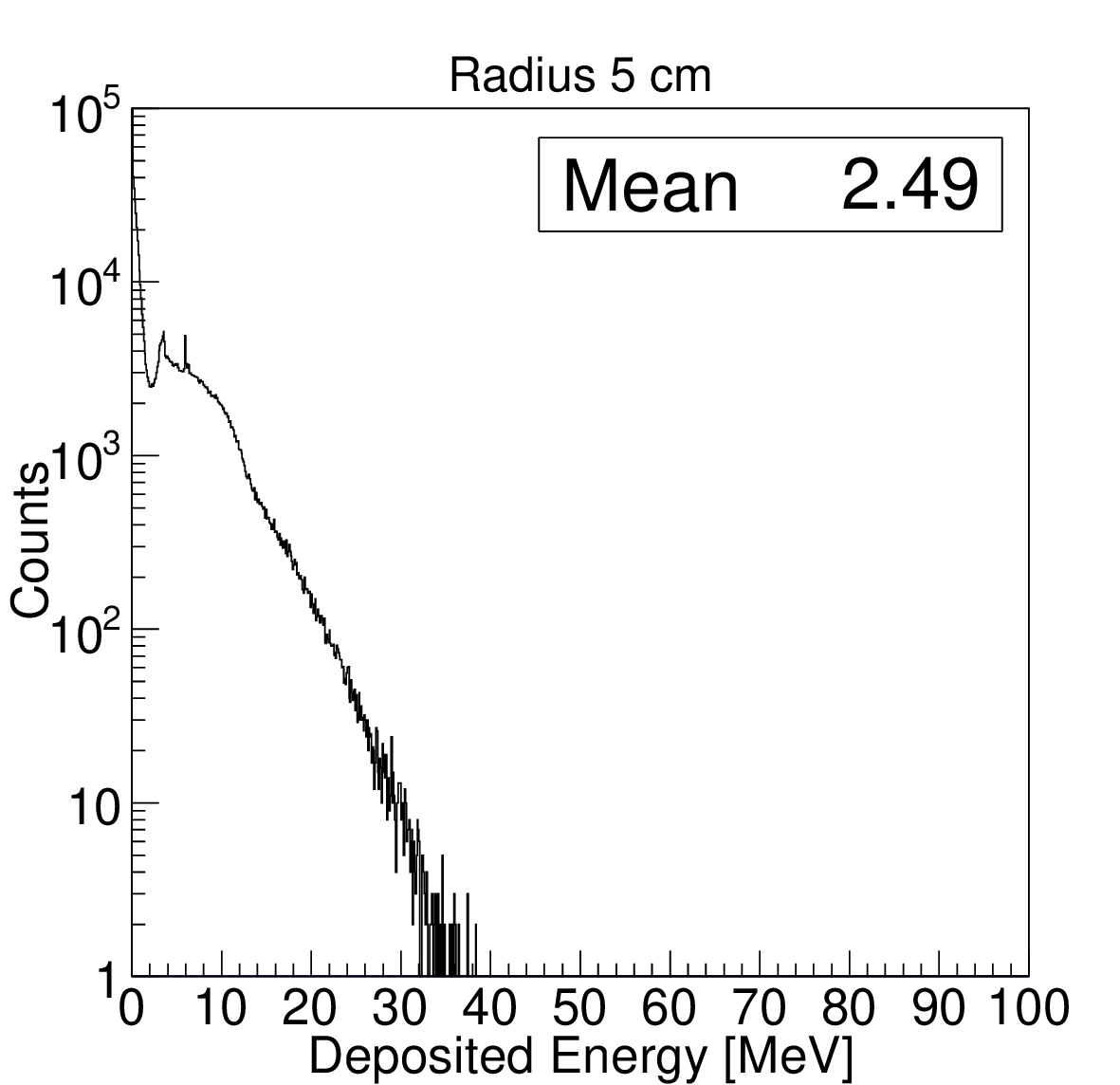} 
	\end{minipage}
		\hspace{0.5cm} 
	\begin{minipage}[t]{0.45\linewidth}
		\includegraphics[width= 0.90\textwidth]{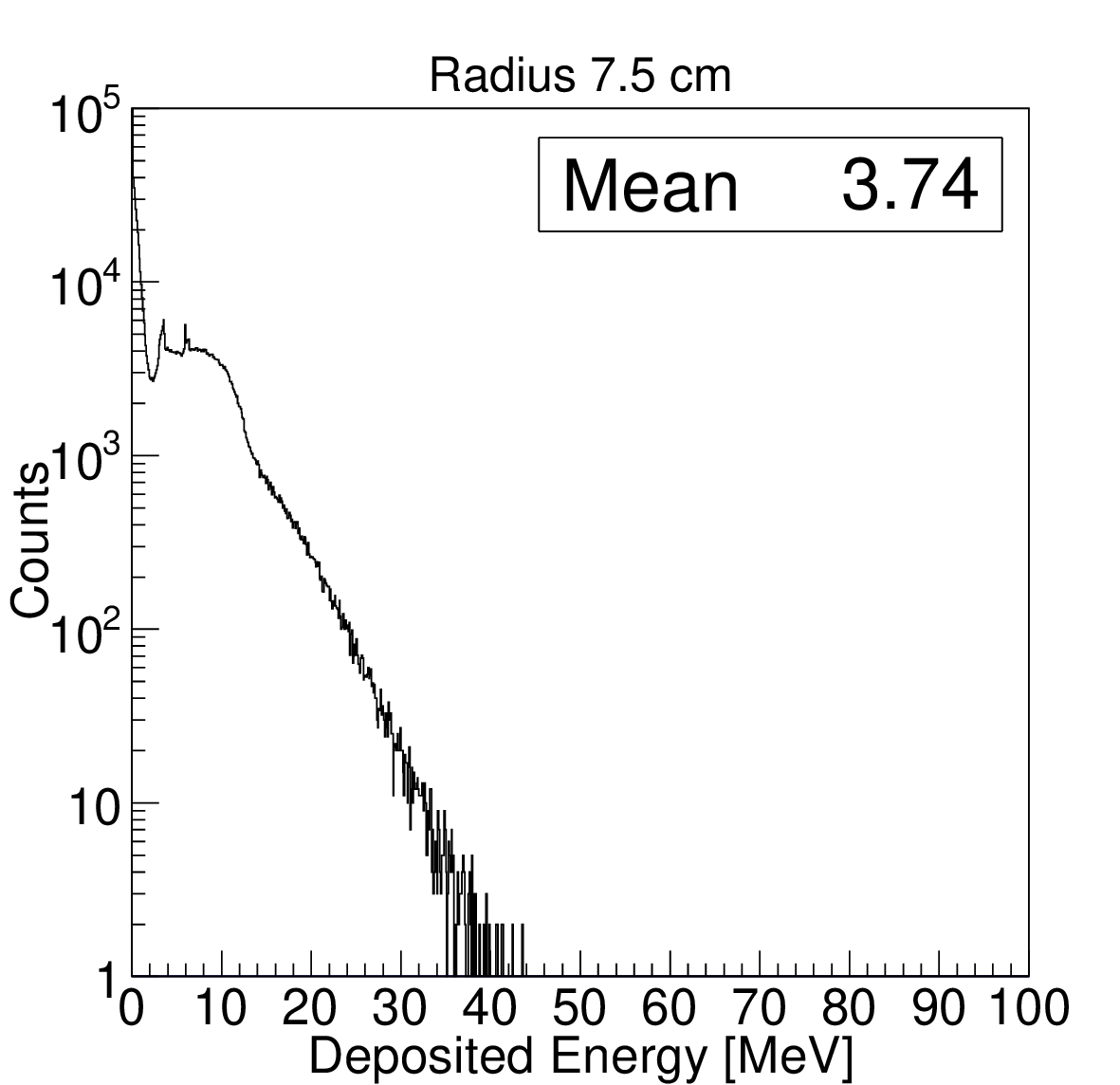} 
	\end{minipage}
\end{figure*}

\begin{figure*}[p!]
\centering
	\begin{minipage}[t]{0.45\linewidth} 
		\centering
		 \includegraphics[width= 0.90\textwidth]{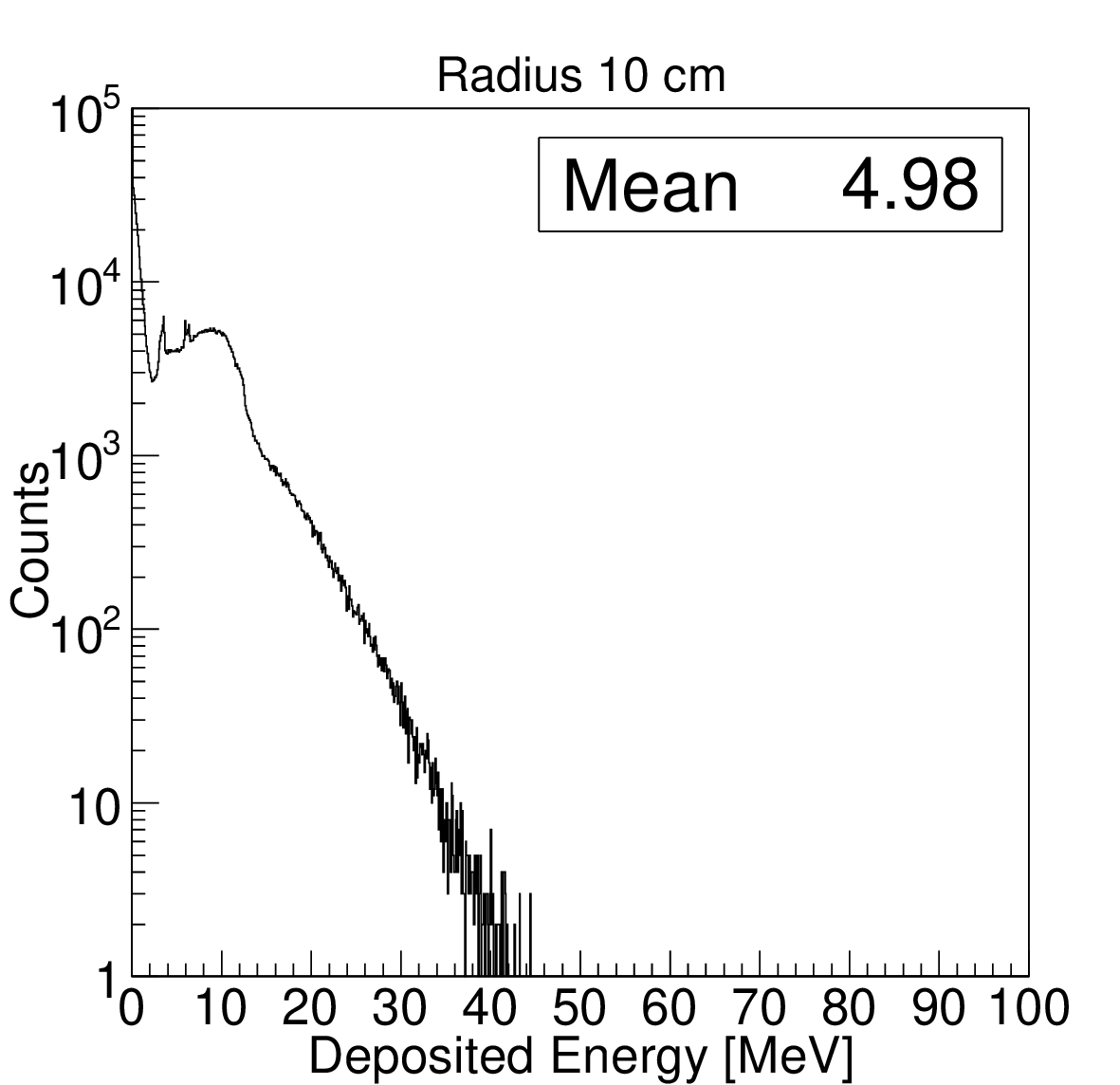} 
	\end{minipage}
		\hspace{0.5cm} 
	\begin{minipage}[t]{0.45\linewidth}
		\includegraphics[width= 0.90\textwidth]{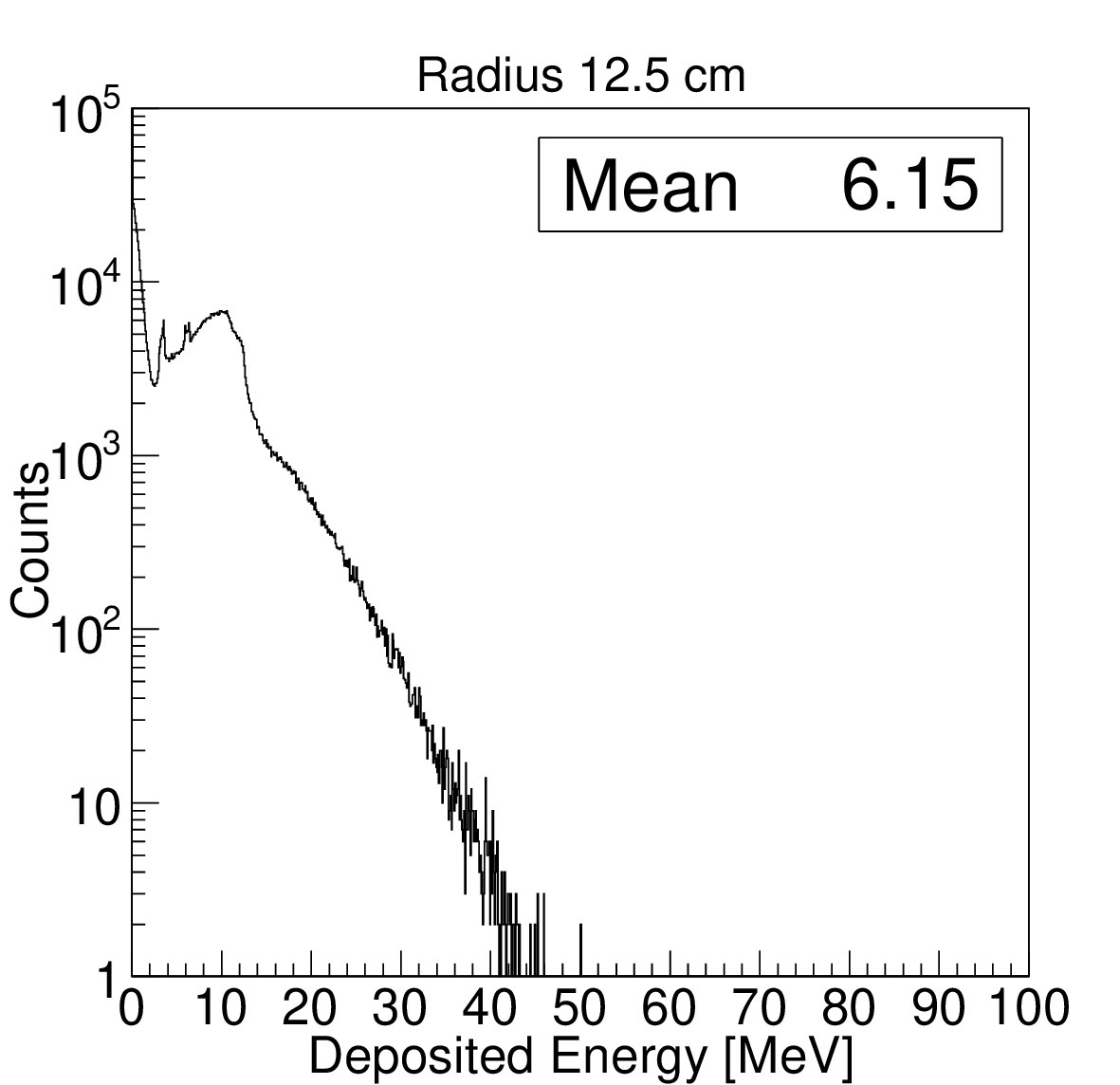} 		 
	\end{minipage}
\end{figure*}

\begin{figure*}[p!]
\centering
	\begin{minipage}[t]{0.45\linewidth} 
		\centering
		 \includegraphics[width= 0.90\textwidth]{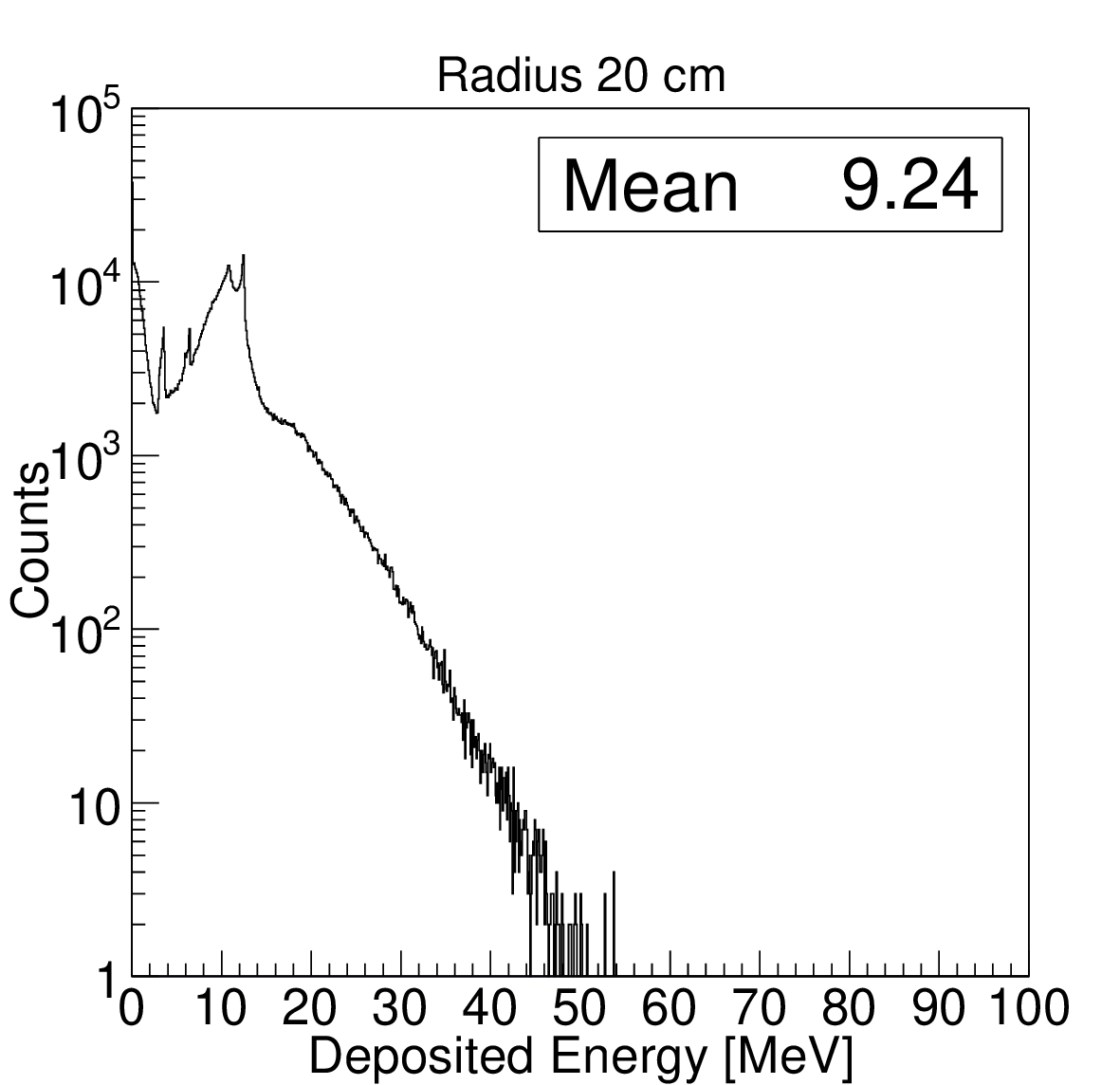} 
	\end{minipage}
		\hspace{0.5cm} 
	\begin{minipage}[t]{0.45\linewidth}
		\includegraphics[width= 0.9\textwidth]{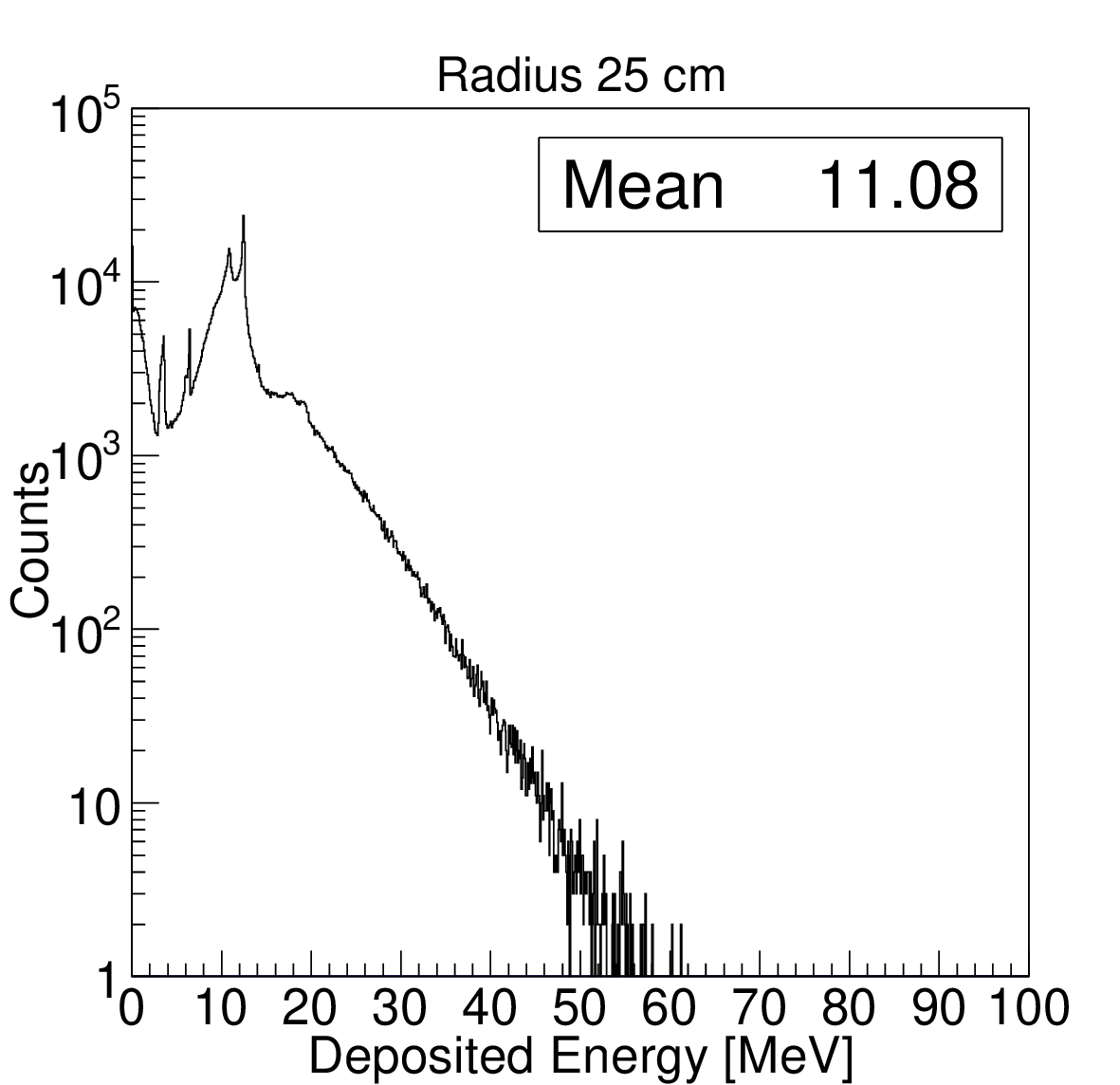} 	 
	\end{minipage}
\end{figure*}

\begin{figure*}[p!]
\centering
	\begin{minipage}[t]{0.45\linewidth} 
		\centering
		 \includegraphics[width= 0.9\textwidth]{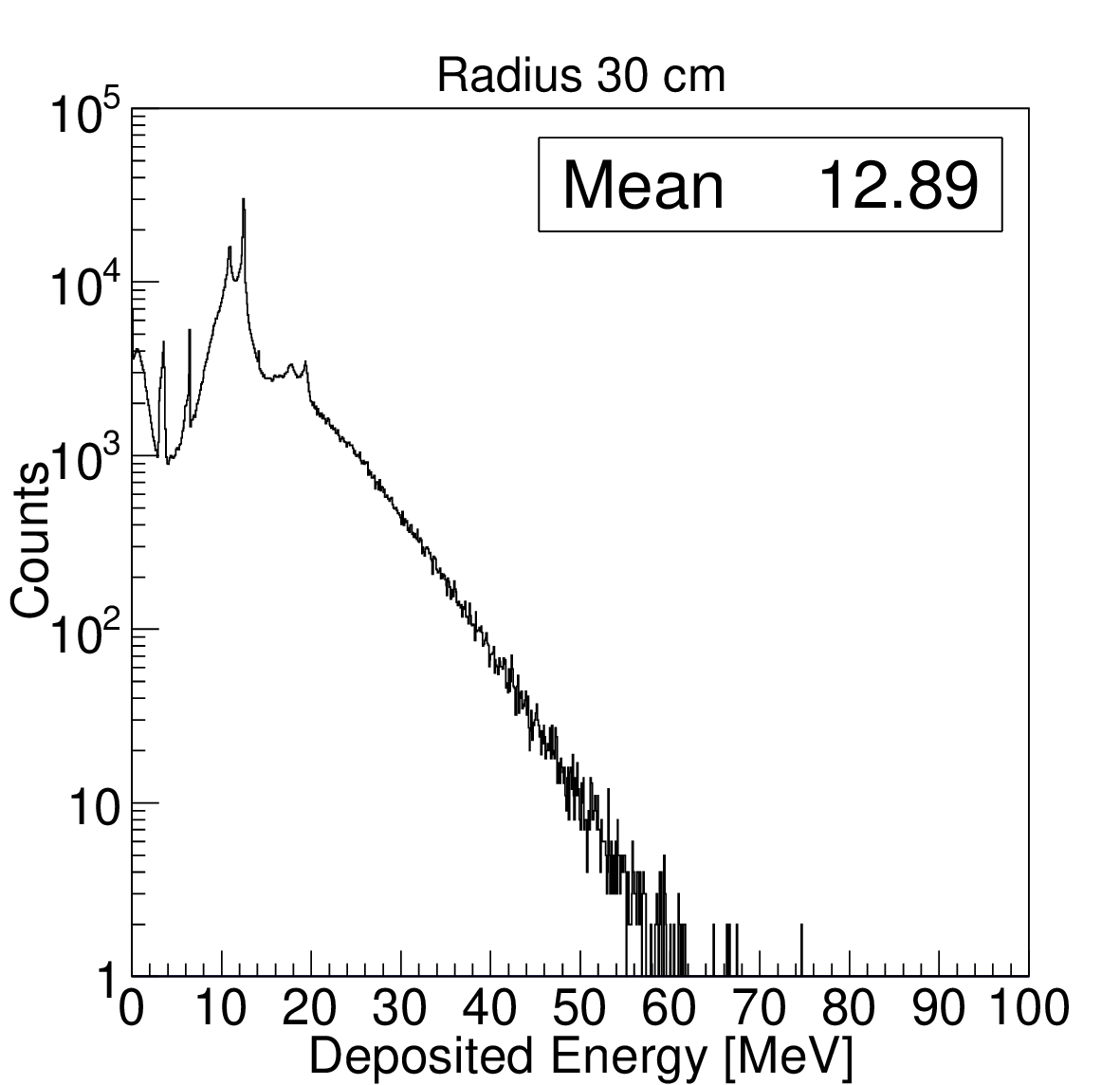} 
	\end{minipage}
		\hspace{0.5cm} 
	\begin{minipage}[t]{0.45\linewidth}
		\includegraphics[width= 0.9\textwidth]{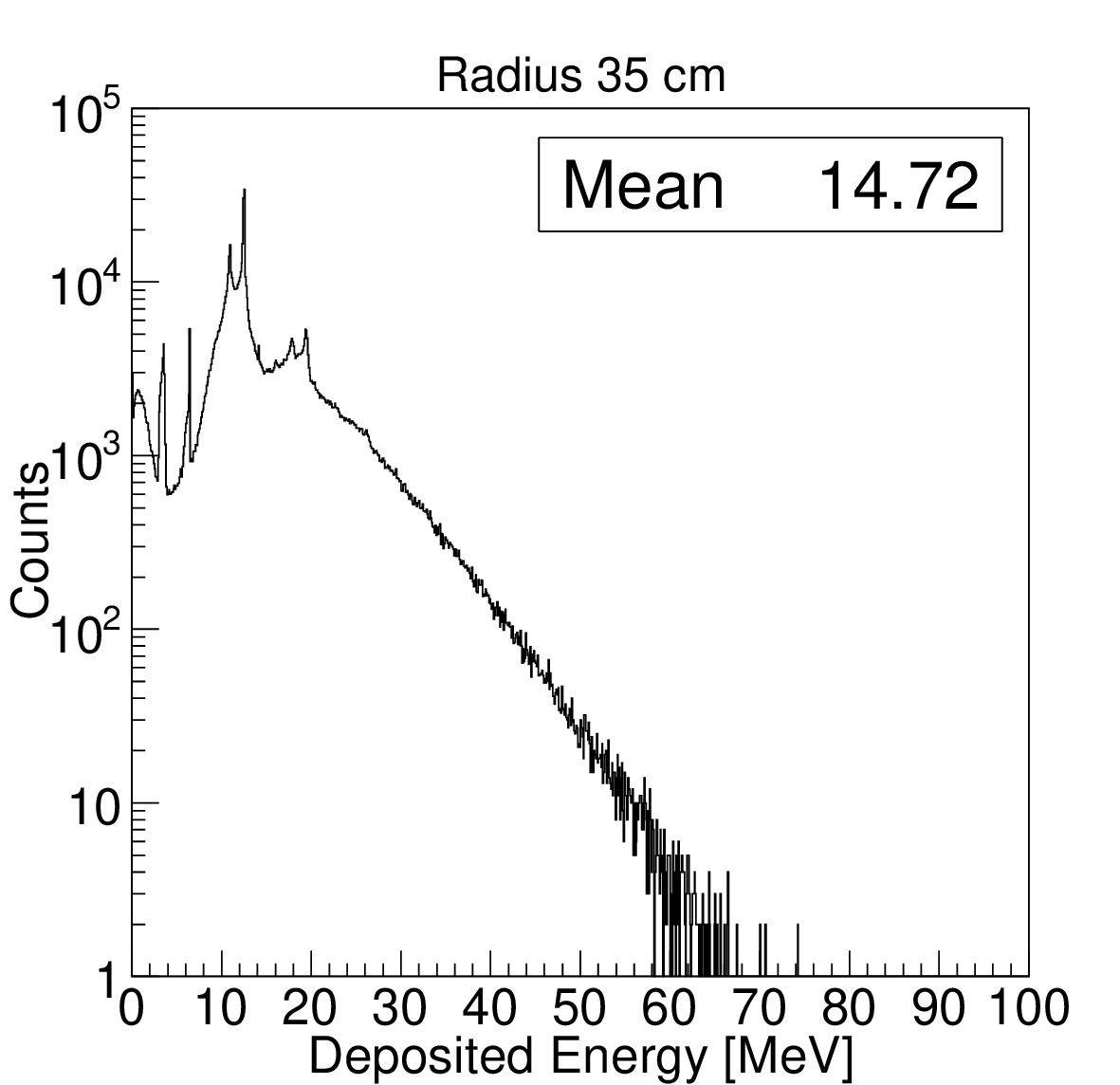} 	 
	\end{minipage}
\end{figure*}

\begin{thebibliography}{9}

\bibitem{GEANT4}
S.~Agostinelli~et~al.,
\emph{Geant4-a simulation toolkit},
http://dx.doi.org/10.1016/S0168-9002(03)01368-8{\emph{Nucl. Instr. Meth. A} {\textbf{506},  (2003)}}.

\bibitem{GEANT4models}
Dennis~H.~Wright~et~al.,
\emph{Low and High Energy Modeling in Geant4}
AIP Conf.Proc.  {\textbf{896 (2007)}} 11-20 SLAC-REPRINT-2007-197 

\end{thebibliography}
\end{document}